\documentclass[aps,prx,twocolumn,superscriptaddress,nofootinbib]{revtex4-2}
\usepackage[T1]{fontenc}
\usepackage{amsmath,amssymb,bm,graphicx}
\usepackage{mathtools}
\usepackage[colorlinks=true,citecolor=blue,urlcolor=blue,linkcolor=blue]{hyperref}
\usepackage{xcolor}
\usepackage[version=4]{mhchem}
\usepackage{siunitx}
\usepackage{braket}
\usepackage{booktabs}
\usepackage{CJKutf8}

\newcommand{\MoTe}{\ensuremath{\mathrm{MoTe}_2}}
\newcommand{\rB}{\ensuremath{\mathbf r}}

\DeclareSIUnit\angstrom{\text {Å}}

\newcommand{\Z}{\mathbb{Z}}

\newcommand{\tr}{\operatorname{Tr}}

\newcommand{\n}[1]{\left| #1 \right|}
\newcommand{\st}[1]{\left\{#1\right\}}

\DeclarePairedDelimiter\ceil{\lceil}{\rceil}
\DeclarePairedDelimiter\floor{\lfloor}{\rfloor}
\let\oldv\v
\renewcommand{\v}[1]{\boldsymbol{#1}}

\def\pbra#1{\mathinner{(#1|}}
\def\pket#1{\mathinner{|#1)}}
\def\pbraket#1{\mathinner{(#1)}}

\usepackage{tikz}
\newlength{\mpsaxis}
\AtBeginDocument{\setbox0=\hbox{$\global\mpsaxis=\fontdimen22\textfont2$}}
\tikzset{
  mpspic/.style = {baseline={([yshift=-\mpsaxis]mpsc)}, line width=0.5pt,
                   inner sep=0pt, outer sep=0pt},
  mps/.style    = {draw, line width=0.5pt, rounded corners=1.5pt, fill=white,
                   minimum width=6mm, minimum height=6mm, inner sep=1pt},
  mpsop/.style  = {draw, line width=0.5pt, fill=white,
                   minimum width=6mm, minimum height=6mm, inner sep=1pt},
  mpsleg/.style = {line width=0.5pt},
}

\usepackage[dvipsnames]{xcolor}
\definecolor{pal0}{rgb}{0.8941, 0.102 , 0.1098}
\definecolor{pal1}{rgb}{0.2157, 0.4941, 0.7216}
\definecolor{pal2}{rgb}{0.302 , 0.6863, 0.2902}
\definecolor{pal3}{rgb}{0.5961, 0.3059, 0.6392}
\definecolor{pal4}{rgb}{1.    , 0.498 , 0.    }

\newcommand{\addcite}[1]{\textsf{\color{red!80!black}[\#]}}

\newcommand{\PRLsec}[1]{\emph{#1---}}

\usepackage[normalem]{ulem}

\begin{document}

\title{Repulsion-Driven Topological Superconductivity in Twisted MoTe$_2$}

\author{Tianle Wang}
\affiliation{Department of Physics, University of California, Berkeley, CA 94720, USA \looseness=-2}
\affiliation{Department of Physics, Harvard University, Cambridge, MA 02138, USA \looseness=-2}

\author{Tomohiro Soejima (\begin{CJK*}{UTF8}{bsmi}副島智大\end{CJK*})}
\affiliation{Center for Computational Quantum Physics, Flatiron Institute,
162 5th Avenue, New York, NY 10010, USA \looseness=-2}
\affiliation{Center for Quantum Phenomena, Department of Physics,
New York University, 726 Broadway, New York, NY, 10003, USA \looseness=-2}

\author{Norman Y. Yao}
\affiliation{Department of Physics, Harvard University, Cambridge, MA 02138, USA \looseness=-2}
\affiliation{Harvard-MIT Center for Ultracold Atoms, Cambridge, Massachusetts 02138, USA \looseness=-2}

\author{Ashvin Vishwanath}
\affiliation{Department of Physics, Harvard University, Cambridge, MA 02138, USA \looseness=-2}

\author{Michael P. Zaletel}
\affiliation{Department of Physics, University of California, Berkeley, CA 94720, USA \looseness=-2}
\affiliation{Material Science Division, Lawrence Berkeley National Laboratory, Berkeley, California 94720, USA \looseness=-2}

\author{Taige Wang}
\email{taige\_wang@fas.harvard.edu}
\affiliation{Department of Physics, Harvard University, Cambridge, MA 02138, USA \looseness=-2}
\affiliation{Materials Research Laboratory, Massachusetts Institute of Technology, Cambridge, MA 02139, USA \looseness=-2}

\author{Daniel E. Parker}
\email{danielericparker@ucsd.edu}
\affiliation{Department of Physics, University of California at San Diego, La Jolla, California 92093, USA \looseness=-2}

\date{\today}

\begin{abstract}
We show that a microscopic model of twisted \ce{MoTe2} with screened Coulomb interactions produces topological superconductivity.
Using the density matrix renormalization group (DMRG) on infinite cylinders, we find that the narrow Chern band of \ce{tMoTe2} has Cooper pair correlations that strengthen with cylinder circumference, consistent with two dimensional superconductivity. The superconductor has chiral $f$-wave pairing around the valley center, and its topology produces chiral Majorana edge modes and non-Abelian vortex cores.
At zero displacement field, the superconductor spans fillings $-0.78\lesssim\nu\lesssim-0.72$, close to the experimental range, and extends in a dome to finite displacement field.
We then predict how its stiffness evolves with the filling and the tunable displacement field. 
The interplay with screening length is also examined.
Nearby, we find a fractional Chern insulator, a topological charge density wave, a metal, and a charge-ordered state. The last is a candidate for the reentrant integer Hall regime in experiments. These results establish a route from repulsive interactions in microscopic models of moir\'e materials to chiral topological superconductors with chiral Majorana edge modes.
\end{abstract}

\maketitle

The potential emergence of superconductivity from purely repulsive interactions has been extensively studied in non-topological settings, most prominently the intermediate-strength Hubbard model~\cite{Arovas2022Hubbard}. Surprisingly,  numerically demonstrating that systems such as the (extended) Hubbard model exhibit all-electronic superconductivity remains a challenge.
Two dimensional materials offer a distinct route to all-electronic superconductivity, as their flat bands feature non-trivial band geometry, which, together with orbital ferromagnetism, can result in Chern bands.s
In twisted \ce{MoTe2}~\cite{Park2023FractionalQAH,XuFan2023QAH} and rhombohedral graphene~\cite{lu2024fractional}, fractional fillings of these Chern bands gives rise to fractional quantum Hall effects at zero magnetic field.  
Can the interplay of repulsive interactions, band geometry, and fractionalization give rise to all-electronic superconductivity?
Recent experiments in twisted bilayer \ce{MoTe2} (\ce{tMoTe2})~\cite{XuFan2025Superconductivity} report superconductivity near an $\nu = -2/3$ fractional Chern insulating state, bringing this question into a concrete microscopic setting.

\begin{figure}[t]
\centering
\includegraphics[width=\linewidth]{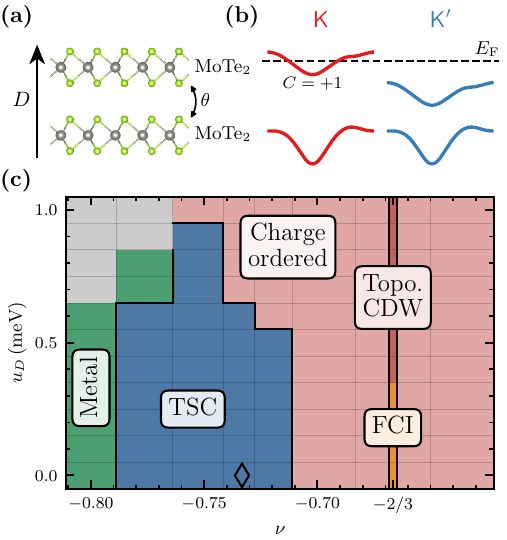}
\caption{
(a) Schematic side view of \ce{tMoTe2} at twist angle $\theta$ and displacement field $D$.
(b) Band structure at $\nu=-0.75$, where the system is valley polarized, yielding a circular Fermi surface centered at the $\gamma$ point.
(c) Inferred iDMRG phase diagram for a microscopic model of \ce{tMoTe2} with strongly screened Coulomb interactions at screening length $d_s =\SI{1.5}{nm}$. A topological superconductor (TSC) competes with a metal and charge-ordered (CO) phase. At nearby filling $\nu=-2/3$, a fractional Chern insulator (FCI) transitions to a $C=1$ charge density wave (Topo. CDW). Gray regions indicate poor DMRG
convergence. Methods and parameters are given in the text.
}
\label{fig:1_overview_phase_diagram}
\end{figure}

The experimental phase diagram of twisted bilayer \ce{MoTe2} ( Fig.~\ref{fig:1_overview_phase_diagram}a) in the doping $\nu$ and displacement field $D$ plane is replete with strongly correlated and topological phases. Doping holes into the two narrow (``flat'') Chern bands near $\theta=3.89^\circ$ (Fig.~\ref{fig:1_overview_phase_diagram}b) produces a ferromagnetic Chern insulator at $\nu=-1$ with $\sigma_{xy}=-e^2/h$~\cite{Cai2023FQAH,Zeng2023ThermodynamicFCI,Park2023FractionalQAH,XuFan2023QAH}, part of a broader ferromagnetic regime at $\nu \gtrsim -1.3$. Fractional Chern insulators (FCIs)~\cite{tang2011high,sheng2011fractional,Neupert2011FCI,Regnault2011FCI,sun2011nearly,Li2021SpontaneousFCI,liu2022recent,lu2024fractional}
are observed at $\nu=-2/3$ and other fractions, the first realization of the phase at zero magnetic field~\cite{Cai2023FQAH,Park2023FractionalQAH,XuFan2023QAH,Zeng2023ThermodynamicFCI}. In certain samples, doping additional holes on top of the FCI leads to, in sequence: a re-entrant integer quantum anomalous Hall (RIQAH) phase, a resistance peak at $\nu \simeq -0.7$ and --- finally and most strikingly --- a superconducting dome centered at $\nu\simeq -0.73$~\cite{XuFan2025Superconductivity}. (Thus far, superconductivity is only observed in a few samples, all near $3.9^\circ$.)
The superconductor is unconventional, as ferromagnetism prohibits $s$-wave pairing, while the nearby FCI suggests strong interactions are essential.

The top valence band of \ce{tMoTe2}~\cite{WuFengcheng2019TMD,devakul2021magic,crepel2023anomalous} is a Chern band close to the ``ideal'' limit~\cite{wang2021exact,ledwith2023vortexability,DongJunkai2023CFL} yet---unlike the lowest Landau level---its finite bandwidth allows for mobile charge carriers.
Theorists have proposed a range of superconducting mechanisms in this setting, ranging from weak-coupling Kohn--Luttinger pairing enhanced by Chern band form factors and quantum geometry~\cite{Kohn1965Pairing,Shavit2025Geometry,Jahin2026Geometry,Zou2026KohnLuttinger,XuCheng2025ChiralSC,ChenYinqi2026FiniteMomentum,MayMann2026PairingTopology,Guerci2025ChiralSC,DongZhiyu2026Pairing}, to strong-coupling anyon superconductivity motivated by the nearby FCI~\cite{Laughlin1988FractionalStatistics,Laughlin1988HallSC,ChenYiHong1989AnyonSC,LeeDungHai1991AnyonSC,Shi2025Doping,Divic2025AnyonSC,WangTaige2025ChiralSC,Nosov2026AnyonSC,Pichler2026AnyonSC,ShiSenthil2025NonAbelianSC,ZhangYahui2026HolonSC,FanVishwanathWang2026,WangTaige2026TopologicalSC}. Recent numerics have examined idealized models motivated by \ce{tMoTe2}: Ref.~\cite{WangTaige2025ChiralSC} used the density matrix renormalization group (DMRG) to study a model with a periodic potential but uniform quantum geometry, while Ref.~\cite{Guerci2025ChiralSC} used exact diagonalization to study a model with non-uniform quantum geometry but no periodic potential. Chiral $f$-wave superconductivity was found in both cases. However, microscopic models of \ce{tMoTe2} are considerably more complex, due to their periodic potential and their non-uniform quantum geometry. Whether repulsive interactions produce superconductivity in such microscopic models of \ce{tMoTe2} is unknown.

Here we study a microscopic model of \ce{tMoTe2} with gate-screened Coulomb interactions using DMRG on an infinite cylinder. Our main findings are fivefold:
(I).~We find superconducting correlations at all accessible cylinder circumferences, consistent with a robust superconducting phase in the 2d limit.
(II).~The superconductor has chiral $f-if$ pairing and an odd number of chiral Majorana edge modes, making it a topological superconductor (TSC).
(III).~Its single-electron gap is $\Delta_{1e}<\SI{0.5}{meV}$, with zero-temperature superfluid stiffness $\rho_s\sim\SI{30}{\micro\electronvolt}$, giving an upper bound $T_c\lesssim\SI{500}{mK}$.
(IV).~The phase diagram contains: a superconducting dome at $-0.78\lesssim\nu\lesssim-0.72$ at zero displacement field; a metal; competition between FCIs and a topological CDW at $\nu=-2/3$; and an intervening charge ordered (CO) phase.
(V).~Superconductivity is favored by shorter-ranged interactions but persists up to effective screening lengths of $d_s \sim\SI{5}{nm}$.

\begin{figure*}[t]
    \centering
    \includegraphics[width=\linewidth]{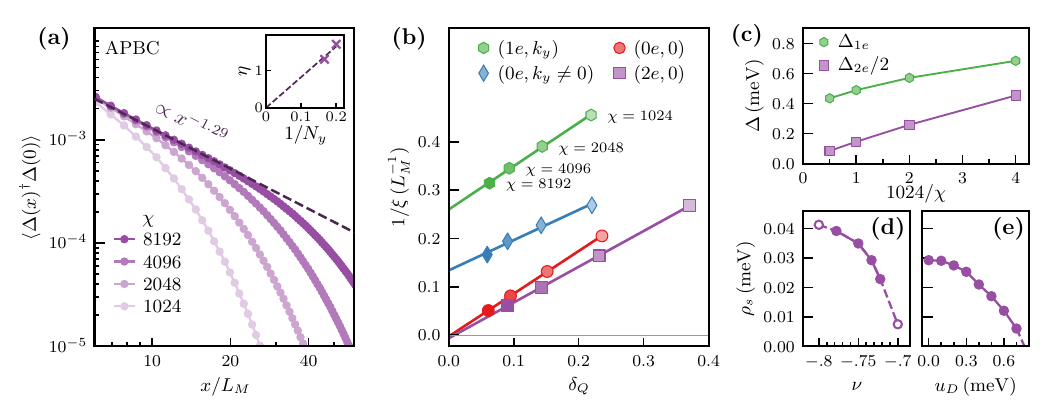}
    \caption{
(a) Cooper-pair correlations at $\nu=-0.73$ and $u_D=0$ (diamond in Fig.~\ref{fig:1_overview_phase_diagram}c), approaching $x^{-\eta_{\mathrm{SC}}}$ with $\eta_{\mathrm{SC}}\approx1.29$. Inset: $\eta_{\mathrm{SC}}$ versus $N_y$, showing the expected $1/N_y$ scaling toward a two-dimensional superconductor. The dashed line is a guide to the eye.
(b) Inverse correlation lengths $1/\xi_Q$ in different charge and momentum sectors versus the finite-entanglement scaling variable $\delta_Q$, where $\delta_Q\to0$ as $\chi\to\infty$. The $(2e,0)$ and $(0e,0)$ sectors extrapolate to zero, while the $(1e,k_y)$ and $(0e,k_y\neq0)$ sectors remain finite.
(c) Charge-$2e$ and single-electron excitation gaps versus $1/\chi$.
(d) Charge stiffness $\rho_s$ versus filling at $u_D=0$. Empty points are outside the SC phase. 
(e) Charge stiffness versus $u_D$ at $\nu=-0.73$.
Parameters: $d_s=\SI{1.5}{nm}$, $\varepsilon=8$, APBC, and $N_y=6$ unless noted.
}
    \label{fig:2_SC_evidence}
\end{figure*}

\PRLsec{Model \& Methods}
We study \ce{tMoTe2} at twist angle $\theta=3.89^\circ$ and hole dopings $-0.8\leq\nu\leq-0.63$. Experiments find strong valley polarization throughout this doping range~\cite{Cai2023FQAH,Zeng2023ThermodynamicFCI,Park2023FractionalQAH,XuFan2023QAH,Redekop2024MagneticImaging,XuFan2025Superconductivity}, so we retain only the $K$ valley (spin is locked to valley). The single-particle Hamiltonian follows a standard continuum model for twisted transition-metal dichalcogenide homobilayers~\cite{WuFengcheng2019TMD} that acts in the layer basis $\ell=(t,b)$ by
\begin{equation}
\label{eq:continuum_Hamiltonian}
\hat h_K(\rB)=
\begin{bmatrix}
h_t(\rB)+u_D & T(\rB)\\
T^\dagger(\rB) & h_b(\rB)-u_D
\end{bmatrix}.
\end{equation}
Here $h_\ell$ contains the monolayer valence-band dispersion and moir\'e-periodic intralayer potential, and $T(\rB)$ describes the moir\'e interlayer tunneling. App.~\ref{app:microscopic_model} gives their precise forms and all other model details. The opposite layer potentials $\pm u_D$ are produced by a perpendicular displacement field $D$. We use the first-harmonic continuum parameters obtained from first-principles calculations in Ref.~\cite{WangChong2024FCI}. At $u_D=0$, the first-harmonic model has $C_{3z}$ and $C_{2y}\mathcal T$ symmetry, plus a pseudoinversion symmetry $\mathcal{I}$ that flips layers and sends $\v{k} \to -\v{k}$ within each valley~\cite{Yu2024FCI,Kolar2024Hofstadter}. A finite displacement field preserves $C_{3z}$ but breaks both $C_{2y}\mathcal T$ and $\mathcal{I}$.

Diagonalizing $\hat h_K$ gives the dispersion $\epsilon_{\v{k}}$ and Bloch eigenstates $\ket{\psi_{\v{k}}}=e^{i\v{k}\cdot\rB}\ket{u_{\v{k}}}$ of the top valence band, which has Chern number $C=+1$ in the $K$ valley. Let $\hat c^\dagger_{\v{k}}$ create the Bloch state $\ket{\psi_{\v{k}}}$. Projecting screened Coulomb interactions into this active band gives
\begin{equation}
\label{eq:projectedH}
H=
\sum_{\v{k}}\epsilon_{\v{k}}\hat c^\dagger_{\v{k}}\hat c_{\v{k}}
+
\frac{1}{2A}\sum_{\v{q}}V_q:
\hat\rho_{\v{q}}\hat\rho_{-\v{q}}:,
\end{equation}
where $A$ is the sample area and $\hat\rho_{\v{q}}=\sum_{\v{k}}\hat c^\dagger_{\v{k}}\braket{u_{\v{k}}|u_{\v{k}+\v{q}}}\hat c_{\v{k}+\v{q}}$ is the projected density operator at momentum $\v{q}$. The interaction is normal ordered with respect to charge neutrality. We take dual gate screened Coulomb interactions $V_q=e^2\tanh(qd_s)/(2\epsilon_0\epsilon_r q)$~\cite{Soejima2020DMRG}, where $\epsilon_r$ is a relative dielectric constant and $d_s$ is an \textit{effective} gate distance that sets the screening length\footnote{
Generically, $d_s$ is related to the geometrical gate distance by a factor determined from the details of the anisotropic dielectric environment, whose value in \ce{tMoTe2} devices is unknown but likely $O(1)$.
}.

We solve Eq.~\eqref{eq:projectedH} using the infinite density matrix renormalization group (iDMRG)~\cite{White1992DMRG,Schollwock2011DMRG,Vanderstraeten2019Tangent} on the infinite cylinder geometry using the TeNPy library~\cite{Hauschild2024TeNPy}. We use a hybrid-Wannier basis and matrix product operator compression techniques developed by some of us~\cite{Parker2020MPO,Soejima2020DMRG}, and previously applied to continuum moir\'e Hamiltonians~\cite{Parker2026FieldTunedFCI,WangTianle2023DMRG,DongJunkai2024HallCrystals} including twisted \ce{MoTe2}~\cite{DongJunkai2023CFL}. We use cylinders with circumference $N_y L_M$, where $N_y=4$--$7$ is the number of moir\'e unit cells around the cylinder and $L_M$ is the moir\'e lattice constant. Equivalently, the allowed transverse momenta form $L_y$ evenly spaced ``wires'' through the Brillouin zone at momenuta $k_y=\frac{2\pi}{N_y L_M}(n_y+\Phi_y/2\pi)$, where $n_y$ is an integer and $\Phi_y$ is the flux threaded through the cylinder. Periodic and anti-periodic boundary conditions (PBC, APBC) correspond to $\Phi_y = 0$ and $\Phi_y=\pi$, respectively. We preserve electric charge $Q_e$, as well as translation symmetry in both the transverse and longitudinal directions. Enlarged MPS unit cells are used when required by the rational filling and/or charge order. We use DMRG bond dimensions up to $\chi=12288$.

\begin{figure}[t]
\begin{center}
\includegraphics[width=\linewidth]{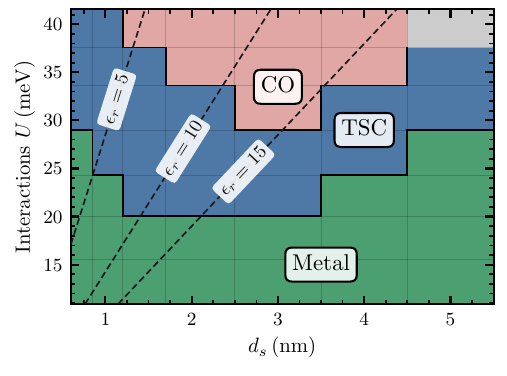}
\end{center}
\caption{
iDMRG phase diagram as a function of screening length $d_s$ and interaction strength $U \equiv nV_{q=0}(d_s,\epsilon_r)/2$ at $\nu=-0.73$ and $u_D=0$. The topological superconductor (TSC) competes with a metal and charge order (CO). Dashed lines indicate $\epsilon_r=5$, $10$, and $15$. Gray regions indicate an unresolved phase. Parameters: $N_y=6$ and APBC.
}
\label{fig:4_gate_eps_phase}
\end{figure}

\PRLsec{Phase diagram.}
Fig.~\ref{fig:1_overview_phase_diagram}(c) shows our central result: a superconducting dome with nearby metallic, charge-ordered, and fractional Chern insulating phases. Here we discuss the phases and their evolution, while their numerical identification is described in the End Matter and App.~\ref{app:phase_identification}. We first focus on strong, short-ranged interactions, $d_s=\SI{1.5}{nm}$ and $\epsilon_r=8$, where the competing phases are clearly resolved. Fig.~\ref{fig:4_gate_eps_phase} shows that superconductivity persists over a broad range of screening lengths and dielectric constants.

Superconductivity appears over $-0.78\lesssim\nu\lesssim-0.72$ at $u_D=0$ and extends to finite $u_D$ in a dome-shaped region. Decreasing filling towards $\nu=-1$ drives a continuous transition to a metal. At $\nu=-0.8$, the metal persists throughout the $u_D$ range studied and has a circular Fermi surface centered at $\gamma$, consistent with the single-particle band minimum shown in Fig.~\ref{fig:1_overview_phase_diagram}(b). At still higher hole density, the metal connects to the valley-polarized Chern insulator at $\nu=-1$ found in previous calculations~\cite{WangTaige2023Magnetism,Qiu2023Magnetism,Abouelkomsan2024Multiferroicity,XuCheng2024Wannier,Yu2024FCI,LiBohao2024Nu1,WangMinxuan2024MagneticField} and experiments~\cite{Cai2023FQAH,Zeng2023ThermodynamicFCI,Park2023FractionalQAH,XuFan2023QAH,Redekop2024MagneticImaging}.

At $\nu=-2/3$, we find an FCI, consistent with previous numerical studies~\cite{Li2021SpontaneousFCI,CrepelFu2023FCI,Reddy2023FQAH,Reddy2023GlobalPhaseDiagram,Qiu2023Magnetism,DongJunkai2023CFL,WangChong2024FCI,MoralesDuran2024MagicFCI,Yu2024FCI}. Increasing $u_D$ drives the FCI into a topological CDW with ordering at $\v{q} = \v{\kappa}$ ($\sqrt{3} \times \sqrt{3}$ order)  and quantized Hall conductivity $\sigma_{xy}=-e^2/h$ (Endmatter).

Doping the superconductor towards $\nu=-2/3$ at any displacement field drives a first-order transition into a ``charge ordered'' (CO) regime whose density $\rho(\v{q})$ is peaked at a wavevector $\v{q} \approx \v{\kappa}$, connecting to the CDW at $\nu=-2/3$. Endmatter and App.~\ref{app:phase_identification} show its order parameter and extent. The precise nature of the charge order is difficult to determine. At larger $u_D$, DMRG on cylinders commensurate with $\sqrt{3}\times\sqrt{3}$ order finds $\rho(\v{q})$ peaked at $\v{\kappa}$ exactly. However, at lower $u_D$ or on incommensurate cylinders, the peak is detuned slightly from $\v{\kappa}$, and the CO may be incommensurate. Away from $\nu=-2/3$, there is no sign of a two-dimensional many-body gap --- indeed a phase which exists over a continuous filling range must be gapless --- however, the DMRG proves hard to converge, and so a definitive identification of the gapless charge excitations remains a challenge.

\PRLsec{Superconductivity.}
We first establish superconductivity at one parameter point, then characterize its stiffness, gap, pairing symmetry and topology below. To identify the ground state, we compute correlation functions of three operators: $\hat{c}^\dagger_{x,k_y}$, which creates an electron in a hybrid Wannier orbital on the ring centered at $x$ along the cylinder with transverse momentum $k_y$; the ring-averaged charge density operator $\hat{n}_x = N_y^{-1}\sum_{k_y} \hat{c}^\dagger_{x,k_y} \hat{c}_{x,k_y}$; and the momentum-resolved Cooper pair operator $\hat{\Delta}_{x,P_y}\sim c_{x,k_y}c_{x,-k_y}$, whose precise form is given in Appendix~\ref{app:phase_identification}.

On a quasi-one-dimensional cylinder, $U(1)$ charge symmetry cannot break spontaneously, so superconductivity manifests as a Luther-Emery liquid (LEL) with quasi-long-range pair correlations~\cite{Luther1974ElectronGas,Emery1976ElectronGas,Gannot2023Cylinders}. At $\nu=-0.73$, $u_D=0$, and $N_y=6$, we find that the $P_y=0$ pair correlations decay algebraically,
\begin{equation}
\label{eq:SC_SC_algebraic_order}
    \left\langle \hat{\Delta}^\dagger_x\hat{\Delta}_0\right\rangle
    \sim x^{-\eta_{\mathrm{SC}}},
\end{equation}
where $\eta_{\mathrm{SC}}$ is the superconducting exponent and $\hat{\Delta}_x\equiv\hat{\Delta}_{x,P_y=0}$. Fig.~\ref{fig:2_SC_evidence}(a) shows that, as the bond dimension increases, the pair correlations approach a power law. A spinless LEL has power-law correlations in the $(Q_e,k_y)=(2e,0)$ and $(0e,0)$ sectors, while correlations in the single-electron $(1e,k_y)$ and finite-momentum neutral $(0e,k_y\neq0)$ sectors decay exponentially~\cite{Luther1974ElectronGas,Emery1976ElectronGas,WangTaige2025ChiralSC}.

To confirm this structure, we perform a scaling analysis of the MPS transfer-matrix~\cite{Vanderstraeten2019Tangent} correlation lengths $\xi_Q$ in each relevant sector. At finite bond dimension $\chi$, algebraic correlations are cut off by the finite MPS correlation length~\cite{Tagliacozzo2008Scaling,Pollmann2009Scaling}. Following the scaling theory of infinite MPS, we extrapolate $1/\xi_Q$ against the transfer-matrix eigenvalue level spacing $\delta_Q$~\cite{Rams2018Extrapolation,Vanhecke2019Scaling,Kuhlenkamp2025Superconductivity} (details in App.~\ref{app:MPS_scaling_theory}). Power-law correlations are signaled by $1/\xi_Q\to0$ as $\delta_Q\to0$, whereas exponentially decaying correlations extrapolate to a finite $1/\xi_Q$. Fig.~\ref{fig:2_SC_evidence}(b) shows that $1/\xi_{(0e,0)}$ and $1/\xi_{(2e,0)}$ extrapolate to zero, while $1/\xi_{(1e,k_y)}$ and $1/\xi_{(0e,k_y\neq0)}$ remain finite. Together, these results identify the ground state as a spinless Luther-Emery liquid at fixed circumference.

We next ask whether pairing strengthens toward the two-dimensional limit. For a two-dimensional superconductor compactified onto a cylinder,
\begin{equation}
\label{eq:eta_Ly_scaling}
\eta_{\mathrm{SC}}(N_y)
=
\frac{\hbar v}{2\pi\rho_s N_y L_M}
\propto \frac{1}{N_y},
\end{equation}
where $\rho_s$ is the two-dimensional superfluid stiffness and $v$ is the phase-mode velocity~\cite{chatterjee2022skyrmion, Gannot2023Cylinders,WangTaige2025ChiralSC}. For $\eta_{\mathrm{SC}}<2$, the superconducting susceptibility diverges at low temperatures as $\chi_{\mathrm{SC}}(T)\sim T^{\eta_{\mathrm{SC}}-2}$~\cite{Arrigoni2004StripedHubbard}. We find $\eta_{\mathrm{SC}}=1.70$ at $N_y=5$ and $1.29$ at $N_y=6$ (see App.~\ref{app:curve_fitting}). The stronger condition $\eta_{\mathrm{SC}}<1$ marks the regime where pairing correlations dominate over competing $2k_F$ density correlations, (a 1d artifact with no 2d limit)~\cite{Luther1974ElectronGas,Emery1976ElectronGas,WangTaige2025ChiralSC}. The decrease from $N_y=5$ to $6$, shown in the inset of Fig.~\ref{fig:2_SC_evidence}(a), is consistent with the $1/N_y$ scaling in Eq.~\eqref{eq:eta_Ly_scaling}, suggesting $\eta_{\mathrm{SC}}<1$ at modestly larger circumference and $\eta_{\mathrm{SC}}\to0$ in the two-dimensional limit.

\PRLsec{Superconducting transition temperature.}
We now compute the two energy scales that characterize the superconducting state: the charge stiffness $\rho_s$, which controls phase coherence, and the single-electron gap $\Delta_{1e}$, which sets the pair-breaking scale. We first determine $\rho_s$ from the dependence of the ground-state energy density $\mathcal{E}$ on the boundary-condition twist $\delta\Phi_y$. The minimum is at $\Phi_y = \pi$, and expanding to second order gives
\begin{equation}
    \label{eq:SC_stiffness}
    \Delta \mathcal E(\delta\Phi_y)
    =
    \frac{\rho_s}{2}
    \left[
    \frac{2 \, \delta\Phi_y}{N_y L_M}
    \right]^2.
\end{equation}
Fig.~\ref{fig:2_SC_evidence}(e) shows $\rho_s$ as a function of $u_D$, extracted from quadratic fits to Eq.~\eqref{eq:SC_stiffness} (App.~\ref{app:curve_fitting}). Fig.~\ref{fig:2_SC_evidence}(d,e) shows $\rho_s$ vanishes either as $\nu$ approaches $-2/3$ or as $u_D$ increases. Within the superconducting phase, $\rho_s$ is typically a few tens of $\si{\micro\electronvolt}$.

We next estimate the superconducting gap $\Delta_{1e}$ using the plane-wave quasiparticle ansatz for MPS~\cite{Haegeman2012Excitations,haegeman2013elementary,Vanderstraeten2019Tangent}. The quasiparticle gap is obtained from the MPS tangent space at fixed length of five rings. Fig.~\ref{fig:2_SC_evidence}(c) shows that the single-electron gap remains finite, with $\Delta_{1e}\lesssim \SI{0.5}{meV}$, while the charge-$2e$ gap $\Delta_{2e}$ is consistent with zero, as expected for a superconductor with gapped single electrons and gapless Cooper pairs.

These energy scales constrain the superconducting transition in two distinct ways. Phase coherence is lost at the BKT transition, where $k_{\mathrm B}T_{\mathrm{BKT}}=\pi\rho_s(T_{\mathrm{BKT}})/2$~\cite{Nelson1977UniversalJump}, while the single-electron gap $\Delta_{1e}$ sets the scale at which pairing itself is destroyed. The superconducting transition is limited by the minimum of the phase-ordering and pair-breaking scales. For $d_s=\SI{1.5}{nm}$, $\Delta_{1e}\sim\SI{0.5}{meV}$ and $\rho_s\sim\SI{30}{\micro\electronvolt}$ correspond to several kelvins and $\sim\SI{500}{mK}$, respectively. Therefore the transition temperature is limited to $T_c\lesssim\SI{500}{mK}$ at this point.

\begin{figure}[t]
    \centering
    \includegraphics[width=\linewidth]{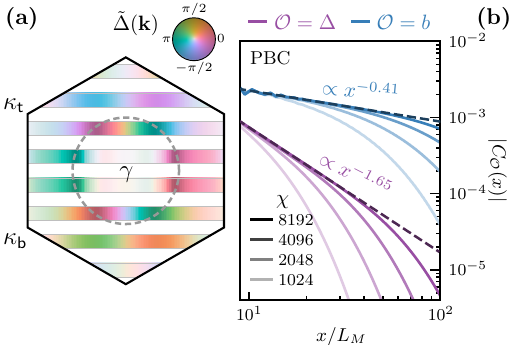}
   \caption{
(a) Proxy pair wavefunction $\widetilde{\Delta}(\v{k})$ in momentum space at $N_y=6$ and $\chi=8192$. Its phase winds as $(k_x-ik_y)^3$ around the Fermi surface (dashed gray circle), indicating chiral $f-if$ pairing.
(b) Cooper-pair ($\mathcal O=\Delta$) and string-order ($\mathcal O=b$) correlations $C_{\mathcal O}(x)$ at $N_y=5$ with periodic boundary conditions (PBC), where $b$ is an electron operator dressed by a nonlocal string. The algebraic string-order correlation identifies a topological superconductor. Parameters in both panels match Fig.~\ref{fig:2_SC_evidence} otherwise.
}
    \label{fig:3_fwave_topological}
\end{figure}

\PRLsec{$f$-wave topological superconductivity.}
We next determine the pairing symmetry and topology of the superconductor. In a two-dimensional superconductor, the pairing symmetry is encoded in the pair wavefunction $\Delta(\v{k})=\braket{\hat c_{\v{k}}\hat c_{-\v{k}}}$, which vanishes identically because our iDMRG preserves charge $U(1)$. We instead measure a proxy pair wavefunction $\widetilde{\Delta}(\v{k})$
from the charge-$2e$ operator with the slowest asymptotic decay along the cylinder (see App.~\ref{app:proxy_pair_wavefunction})~\cite{Sahay2024Superconductivity,Divic2025AnyonSC,WangTaige2025ChiralSC}.
Fig.~\ref{fig:3_fwave_topological}(a) shows that the phase of $\widetilde{\Delta}(\v{k})$ winds by $-6\pi$ around the Fermi contour, corresponding to angular momentum $\ell=-3$ in our convention. This establishes chiral $f$-wave pairing.

If the superconductor were in the weak-pairing limit, its topology would follow from the pair winding. For pairing in a Chern band~\cite{Read2000PairedStates,Qi2010ChiralSC,WangJing2015ChiralSC,WangTaige2025ChiralSC},
\begin{equation}
\label{eq:BdG_Chern_number}
C_{\mathrm{BdG}}=2C+\ell,
\end{equation}
where $C$ is the normal-band Chern number and $\ell$ is the pair winding around the Fermi surface. In this limit, the number of chiral Majorana edge modes is $2c_-=C_{\mathrm{BdG}}$, giving $2c_-=-1$ for $C=1$ and $\ell=-3$. Here $c_-$ is the chiral central charge quantized in half-integer units~\cite{Read2000PairedStates,Kitaev2006Anyons}. Beyond weak pairing, however, pairing symmetry alone does not determine the topology~\cite{Read2000PairedStates,ZhangWangZaletel2026}.

We therefore diagnose the topology directly within number-conserving DMRG~\cite{ZhangWangZaletel2026,Kane2017Pairing}. An odd $2c_-$ implies that a superconducting vortex binds an unpaired Majorana zero mode and has non-Abelian statistics, whereas an even $2c_-$ gives Abelian vortices~\cite{Read2000PairedStates,Kitaev2006Anyons}. On a quasi-one-dimensional cylinder, this even--odd distinction appears as a gapless analogue of one dimensional symmetry protected topological order and can be detected by a nonlocal string operator~\cite{BahriVishwanath2014,ZhangWangZaletel2026},
\begin{equation}
\hat b_x=
\left[\; \prod_{x'<x}(-1)^{\hat N_{x'}}\right]
\hat c_{x,k_y=0},
\end{equation}
where $\hat N_{x'}$ is the total electron number on the transverse ring at $x'$. For zero-momentum pairing, bosonization gives~\cite{ZhangWangZaletel2026}
\begin{equation}
    \big|\braket{\hat{b}_x^\dagger \hat{b}_0}\big| \sim
    \begin{cases}
        e^{-x/\xi_b}, & 2c_- = 0 \; \text{(mod 2) [Abelian]},\\
        |x|^{-\eta_b}, & 2c_- = 1\; \text{(mod 2) [non-Abelian]}
    \end{cases}
\end{equation}
with $\eta_b=\eta_{\mathrm{SC}}/4$. Fig.~\ref{fig:3_fwave_topological}(b) shows algebraic decay with $\eta_b\approx0.41$, while the Cooper-pair exponent $\eta_{\mathrm{SC}}\approx1.65$ satisfies $\eta_{\mathrm{SC}}/4=\eta_b$ within fitting error. This establishes an odd number of chiral Majorana edge modes and hence non-Abelian vortices.

\PRLsec{Screening-controlled superconductivity.} Finally, we examine how the range and strength of the Coulomb interaction controls superconductivity. We parameterize the interaction strength at hole density $n = |\nu| / A_M$ as $U\equiv nV_{q=0}(d_s,\epsilon_r)/2$ where, as above, $d_s$ is an \textit{effective} screening length related to, but not necessarily equal to, the geometrical gate spacing. As shown in Fig.~\ref{fig:4_gate_eps_phase}(a), weak interactions favor a metal, while stronger repulsion generally favors the charge order, with the topological superconductor in between. Lines of constant $\epsilon_r$ indicate representative parameters for moir\'e devices.
Fig.~\ref{fig:4_gate_eps_phase} establishes that superconductivity can reach $d_s\approx\SI{5}{nm}$. However, numerical convergence becomes more difficult at larger $d_s$.

\PRLsec{Discussion.}
We have established repulsion-driven chiral $f$-wave topological superconductivity with non-Abelian vortices in a microscopic continuum model of \MoTe{} with gate-screened Coulomb interactions. The superconductor appears at $-0.78\lesssim\nu\lesssim-0.72$ at zero displacement field and survives at finite $u_D$, showing that it does not rely on inversion symmetry. 
Whether the pairing originates from a Kohn--Luttinger instability, strong-coupling anyon superconductivity, or another mechanism remains an open question. However, the increase of $\rho_s$ with doping away from $\nu=-2/3$ (Fig.~\ref{fig:2_SC_evidence}d) is consistent with doped-FCI scenarios.

Our numerics make several concrete experimental predictions. Increasing hole doping away from $\nu=-2/3$ enhances the superfluid stiffness, while increasing $u_D$ strongly suppresses it even as the single-electron gap remains comparatively robust. Microwave measurements of $\rho_s$~\cite{Tanaka2025Stiffness,Banerjee2025Stiffness} together with tunneling spectroscopy~\cite{Park2026TunnelingGap} can directly test these trends.
Phase-sensitive measurements that account for the spin-valley, finite-momentum, and Bloch-orbital structure of the intravalley pairs can test the predicted chiral $f$-wave pairing~\cite{Xiao2023Josephson,SainzCruz2023Junctions}, while vortex spectroscopy and thermal transport can potentially probe its topology~\cite{Machida2019VortexMajorana,Jack2021MajoranaSTM,Yang2022ThermalMajorana}. Electrostatic screening is another experimentally tunable knob, driving the system from charge order through the TSC to the metal (Fig.~\ref{fig:4_gate_eps_phase}). It can be modulated experimentally by varying the gate distance or dielectric environment~\cite{Barrier2026Screening,Gao2026DielectricScreening}. 

The charge-ordered regime provides a possible microscopic origin of the reentrant Hall state. At $\nu=-2/3$, increasing $u_D$ drives a transition from the FCI to a topological CDW. In experiments, loss of valley polarization (which is beyond the scope of our single-valley calculation) may preempt this transition~\cite{Redekop2024MagneticImaging}. 
Upon doping away from $\nu=-2/3$, we continue to find near-$\sqrt{3}\times\sqrt{3}$ charge order, although DMRG convergence becomes more challenging and its detailed nature is less certain.
In this doped regime, disorder may Anderson localize the residual carriers, exposing the integer Hall response of the topological CDW, thus providing a route to the reentrant integer quantum anomalous Hall (RIQAH) regime~\cite{XuFan2025Superconductivity,Huang2025Streda}.
This charge order regime is distinct from the
translation-symmetric $\Z_3$ orthogonal metal~\cite{Shi2026AnyonFluid,Senthil2026FractionalizedMetals,Han2026OrthogonalMetal}, or coexistence of FCI, charge-ordered, and superconducting domains~\cite{Lotric2026Transition} and thermally stabilized non-Fermi liquid metals~{\cite{ZhangYahui2026HolonSC,FanVishwanathWang2026,WangTaige2026TopologicalSC,zhang2026colorsuperconductorsholonmetals,mehta2026coloringanyonsuperconductivity}}. These scenarios can be distinguished experimentally through imaging and transport. The predicted near-$\sqrt{3}\times\sqrt{3}$ modulation can be directly detected in microscopy~\cite{Thompson2025STM,Liu2026STM}, while the $\Z_3$ orthogonal metal, domain-coexistence, and charge-order scenarios give distinct trajectories in the $(\rho_{xy},\rho_{xx})$ plane~\cite{Senthil2026FractionalizedMetals,Han2026OrthogonalMetal,Lotric2026Transition}.

\PRLsec{Note Added.} During the preparation of this work, Refs.~\cite{OuyangWutMoTe2SC,Fonseca2026Zoology,Pichler2026Attraction} appeared. Ref.~\cite{OuyangWutMoTe2SC} finds $p$-wave rather than $f$-wave superconductivity at similar fillings in a related \ce{tMoTe2} model, while Refs.~\cite{Fonseca2026Zoology,Pichler2026Attraction} find broader families of chiral superconducting states in minimal Chern-band models with different pairing symmetries.

\begin{acknowledgments}
We thank T. Senthil and Zijian Wang for helpful discussions. Our computations were run on the FASRC Cannon cluster supported by the FAS Division of Science Research Computing Group at Harvard University. Tianle Wang acknowledges support from the Schwinger Foundation Postdoctoral Fellowship (JSF-25-09-0004). N.Y.Y. acknowledges support from a Brown Investigator Award. M.P.Z. and Tianle Wang acknowledge support from the U.S. Department of Energy, Office of Science, Office of Basic Energy Sciences, Materials Sciences and Engineering Division under Contract No.~DE-AC02-05CH11231 (Theory of Materials program KC2301). D.E.P. acknowledges support from NSF CAREER Award No.~DMR-2542485. Taige Wang and A.V. acknowledge support from the Simons Collaboration on Ultra-Quantum Matter, a grant from the Simons Foundation (651440, A.V.). Taige Wang acknowledges additional support from the Harvard Quantum Initiative Fellowship. The Flatiron Institute is a division of the Simons Foundation.
\end{acknowledgments}

\section*{End Matter}

\begin{figure}[t]
\begin{center}
\includegraphics[width=\linewidth]{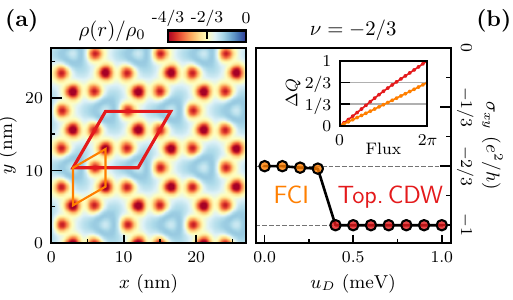}
\end{center}
\caption{\textbf{FCI--CDW transition with displacement field.} 
(a) 
Real-space charge density $\rho(r)$ at $\nu=-0.67$ and $u_D=\SI{0.5}{meV}$. Here $\rho_0=-1/A_{\mathrm{uc}}$ is the density at $\nu=-1$. A $\sqrt{3}\times\sqrt{3}$ pattern is clearly visible.
(b) Hall conductivity $\sigma_{xy}$ versus $u_D$ at $\nu=-2/3$, obtained from charge pumping.
(Inset) Charge pumping from flux through the cylinder at $u_D = 0$ and $u_D = \SI{0.5}{meV}$ respectively. Parameters in both panels match Fig.~\ref{fig:2_SC_evidence} unless stated.
}
\label{fig:5_FCI-CDW}
\end{figure}

\begin{figure}[h!]
\begin{center}
\includegraphics[width=\linewidth]{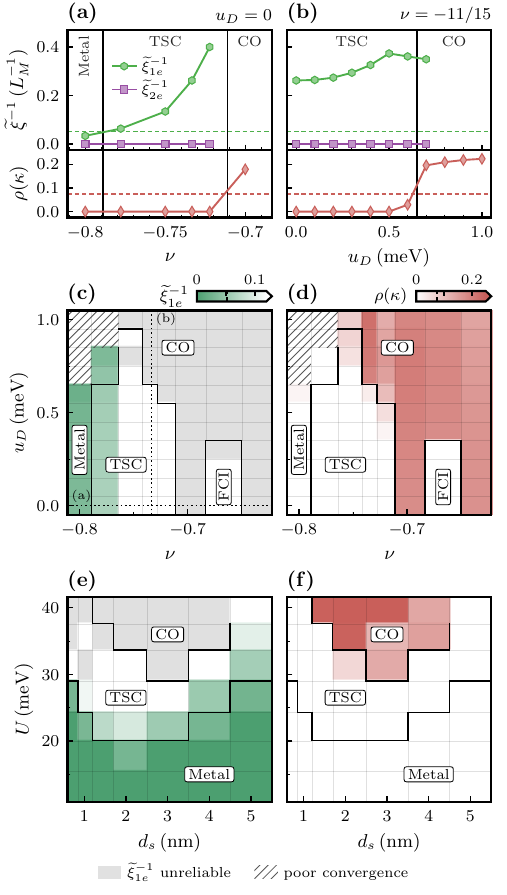}
\end{center}
\caption{
Numerical diagnostics used to identify the phases in Fig.~\ref{fig:1_overview_phase_diagram}(c) and \ref{fig:4_gate_eps_phase}. The metal is identified by the extrapolated inverse single-electron correlation length $(\tilde{\xi}_{1e})^{-1}$, which vanishes in the metal, while charge order is identified by the density-wave order parameter $\rho(\kappa)$ defined in the End Matter. The superconducting region is identified using several diagnostics, with the Cooper-pair exponent shown across the phase diagrams in Fig.~\ref{fig:7_sc_phase_identification}.
(a,b) $(\tilde{\xi}_{1e})^{-1}$, $(\tilde{\xi}_{2e})^{-1}$, and $\rho(\kappa)$ across linecuts of (c), showing the thresholds and phase transitions.
(c,d) $(\tilde{\xi}_{1e})^{-1}$, and $\rho(\kappa)$ in the filling $\nu$ and displacement field $u_D$ phase diagram of Fig.~\ref{fig:1_overview_phase_diagram}(c). (e,f) The same diagnostics in the interaction strength $U$ and screening length $d_s$ phase diagram of Fig.~\ref{fig:4_gate_eps_phase}. Throughout, black lines indicate the inferred phase boundaries. Gray and hatched regions indicate unreliable correlation-length extrapolation and poor DMRG convergence. We measure $\rho(\kappa)$ at $\chi=2048$.}
\label{fig:6_phase_identification}
\end{figure}

\PRLsec{Topological CDW and its transition to the FCI.} To study competition between charge order and topological order at $\nu=-2/3$, we performed DMRG at $N_y=6$ in a larger variational space whose translation symmetry was reduced to permit enlarged unit cells with $\sqrt{3}\times\sqrt{3}$ order. At $u_D=0$ we find the well-studied Laughlin FCI, while at high displacement field we find a state with exponentially decaying correlations in all channels, signaling a many-body gap. Fig.~\ref{fig:5_FCI-CDW}(a) shows this state's real-space density $\rho(\v r)$, where a clear $\sqrt{3}\times\sqrt{3}$ modulation is directly visible. Related commensurate charge-ordered states have been found previously in exact-diagonalization studies of twisted \MoTe{}~\cite{crepel2023anomalous,Reddy2023FQAH,SharmaPengSheng2024,ZaklamaLuoFu2025}, including a study of their real-space density in Ref.~\cite{SharmaPengSheng2024}.

The density forms well-localized ``fidget spinner'' motifs, akin to the Wannier functions in the topological bands of twisted bilayer graphene~\cite{po2018origin}, suggesting that the charge order may retain nontrivial topology. To confirm this, we thread $2\pi$ magnetic flux through the cylinder, and measured the pumped charge $\Delta Q$, which reflects the Hall conductivity $\sigma_{xy} = -(e^2/h) \Delta Q$. Fig.~\ref{fig:5_FCI-CDW}(b, inset) shows the expected  $\sigma_{xy}=-\frac{2}{3}e^2/h$ for the FCI at $u_D=0$ and $\sigma_{xy}=-e^2/h$ in the CDW, confirming it as a topological CDW (TCDW). Charge pumping across a range of displacement fields, Fig.~\ref{fig:5_FCI-CDW}(b), places the FCI/TCDW transition near $u_D\approx\SI{0.45}{meV}$. 

\PRLsec{Charge Order Parameter}
Away from $\nu=-2/3$, charge order persists over a finite range of fillings, manifesting through peaks in the Fourier space charge density $\rho(\v{q})$ away from the moir\'e reciprocal lattice.
As shown in Fig.~\ref{fig:app:phase_identification} (App.~\ref{app:phase_identification}), $\rho(\v q)$ develops pronounced peaks near the six $\v\kappa_j$ points at the corners of moir\'e Brillouin zone, showing that the order remains close to $\sqrt{3}\times\sqrt{3}$ even at incommensurate filling. For all dopings, we observe non-trivial peaks in $\rho(\v{k})$ only in the vicinity of $\v{\kappa}$.

We quantify this charge order by
\begin{equation}
    \rho(\kappa)=\max_{\v q\approx\v\kappa_j}
    \frac{\left|\left\langle\hat\rho_{\v q}\right\rangle\right|}
    {\left|\left\langle\hat\rho_{\v 0}\right\rangle\right|}.
    \label{eq:cdw_order}
\end{equation}
The allowed momenta $\v{q}$ depend on the filling-dependent DMRG unit cell on the cylinder, and do not always include $\v{\kappa}_j$. On $\v{\kappa}$-commensurate cylinders and at larger displacement fields, the peak lies exactly at $\v\kappa_j$, while at lower displacement field or on incommensurate cylinders it can shift away. As noted above, the state should be gapless away from $\nu=-2/3$, but DMRG with expanded unit cells is expensive and suffers from convergence issues, preventing a clear determination of the nature of any gapless charge excitations. We leave a definitive characterization of this phase to future work.

\begin{figure}[t]
\begin{center}
\includegraphics[width=\linewidth]{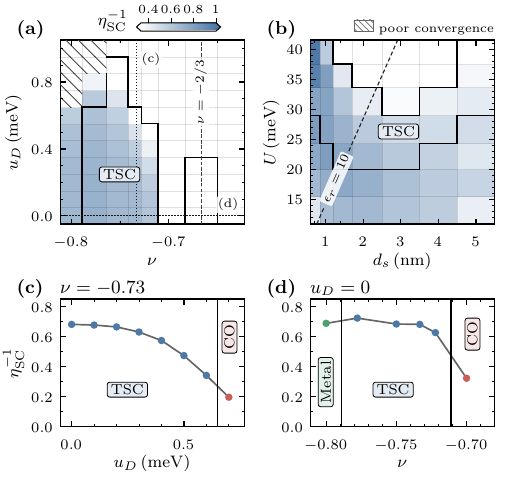}
\end{center}
\caption{Superconducting correlations across the phase diagrams in Fig.~\ref{fig:1_overview_phase_diagram}(c) and \ref{fig:4_gate_eps_phase}. We characterize the strength of Cooper-pair correlations by the inverse superconducting exponent $\eta_{\rm SC}^{-1}$, where $\left\langle \hat{\Delta}^\dagger_x\hat{\Delta}_0\right\rangle\sim x^{-\eta_{\mathrm{SC}}}$. This is non-zero in the metal. The topological superconductor is identified using this together with the additional diagnostics in Fig.~\ref{fig:2_SC_evidence}. (a) $\eta_{\rm SC}^{-1}$ in the filling $\nu$ and displacement field $u_D$ phase diagram of Fig.~\ref{fig:1_overview_phase_diagram}(c). (b) $\eta_{\rm SC}^{-1}$ in the interaction strength $U$ and screening length $d_s$ phase diagram of Fig.~\ref{fig:4_gate_eps_phase}. (c,d) Line cuts through (a), showing $\eta_{\rm SC}^{-1}$ versus $u_D$ at $\nu=-0.73$ and versus $\nu$ at $u_D=0$, respectively. Black lines indicate the inferred phase boundaries. Hatched regions indicate poor DMRG convergence.}
\label{fig:7_sc_phase_identification}
\label{fig:7_sc_phase_identification}
\end{figure}

\PRLsec{Phase identification.} Fig.~\ref{fig:6_phase_identification} summarizes the diagnostics used to distinguish the TSC from its competing metallic and charge-ordered phases in Figs.~\ref{fig:1_overview_phase_diagram}(c) and \ref{fig:4_gate_eps_phase}. In both cases, the TSC is sandwiched between a metal on one side and charge order on the other.

We identify charge order using $\rho(\kappa)$ from Eq.~\eqref{eq:cdw_order}, classifying points with $\rho(\kappa)\geq0.075$ at $\chi=2048$ as charge ordered. To distinguish the metal from the TSC, we use the single-electron transfer-matrix correlation length. Following the finite-entanglement scaling procedure used in the main text and described in detail in App.~\ref{app:phase_identification}, we define $(\tilde{\xi}_Q)^{-1}\equiv\lim_{\chi\to\infty}\xi_Q(\chi)^{-1}$. Gapless charge-$Q$ excitations give $(\tilde{\xi}_Q)^{-1}=0$, while a finite value signals exponentially decaying correlations in that sector. Among points without charge order, we identify the metal by $(\tilde{\xi}_{1e})^{-1}\leq0.045$. 

The two criteria above are necessary but not sufficient to identify a superconductor. For converged states that are non-metallic and non-charge ordered, we
verify superconductivity using the diagnostics established in Fig.~\ref{fig:2_SC_evidence}, including algebraic Cooper-pair correlations, a gapped single-electron sector, and gapless charge-$2e$ correlations. We identify this region with the TSC characterized in Fig.~\ref{fig:2_SC_evidence} and \ref{fig:3_fwave_topological}. Fig.~\ref{fig:6_phase_identification} shows $\rho(\kappa)$ grows upon approaching charge order, while $(\tilde{\xi}_{1e})^{-1}$ decreases toward zero upon approaching the metal. 

\PRLsec{Superconducting correlations.} Fig.~\ref{fig:7_sc_phase_identification} shows the inverse superconducting exponent $\eta_{\rm SC}^{-1}$ across both phase diagrams, defined via $\left\langle \hat{\Delta}^\dagger_x\hat{\Delta}_0\right\rangle\sim x^{-\eta_{\mathrm{SC}}}$ and computed via scaling collapse fit (App.~\ref{app:curve_fitting}). This is expected to be non-zero in a metal, where its value is set by the gapless electron-electron correlations. As discussed in the main text, $\eta_{\rm SC}<2$ gives a divergent superconducting susceptibility, while $\eta_{\rm SC}<1$ marks the regime where Cooper-pair correlations dominate over competing $2k_F$ density correlations. The estimates of $\eta_{\rm SC}$ across the phase diagram are less precise than the detailed finite-entanglement analysis in Fig.~\ref{fig:2_SC_evidence}, so we do not use them to determine the phase boundaries, but they are sufficient to track the evolution of pairing correlations.

Throughout the TSC region, the Cooper-pair correlations remain strong. On entering the charge-ordered region, the fitted $\eta_{\rm SC}$ increases rapidly and the pair correlations become short ranged. Toward the metal, $\eta_{\rm SC}$ instead increases more gradually as the Cooper-pair correlations weaken. Fig.~\ref{fig:7_sc_phase_identification}(c,d) shows these trends along two cuts through panel (a), at fixed $\nu=-0.73$ and fixed $u_D=0$, respectively. The behavior of $\eta_{\rm SC}$ is consistent with the superconducting region identified independently from charge order and the single-electron correlation length.

\let\temp\oldv
\let\oldv\v
\let\v\temp

\bibliographystyle{apsrev4-2}
\bibliography{mote2_refs}

\let\temp\oldv
\let\oldv\v
\let\v\temp

\clearpage

\appendix

\begin{figure*}[t]
    \centering
    \includegraphics[width=\linewidth]{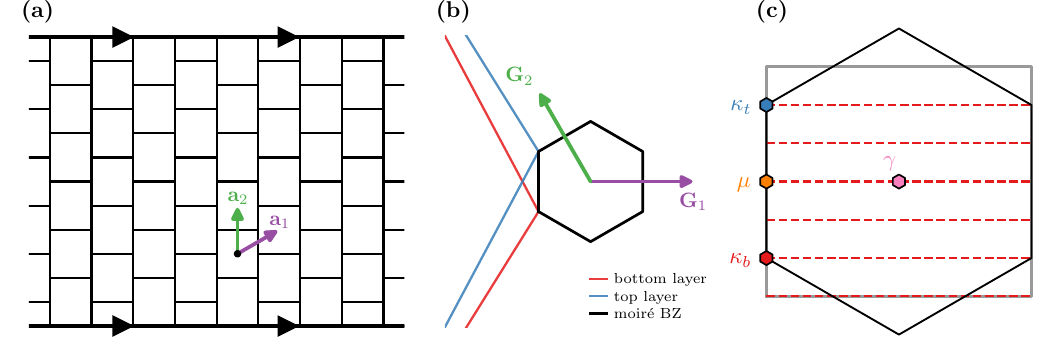}
    \caption{
    (a) Real-space geometry of the infinite cylinder on which DMRG is performed. The rectangular unit cells tile the cylinder in a brick wall pattern, shown here for circumference $L_y=6$. Each vertical column of unit cells forms a ``ring'' of the cylinder of width $L_x$. Here $\v{a}_{1,2}$ generates the moir\'e superlattice. The black arrows indicate that the top and bottom edges are identified. 
    (b) Momentum-space geometry of \ce{tMoTe2}, showing the monolayer Brillouin zones and the moir\'e Brillouin zones at $\theta=3.89^\circ$.
    (c) Moir\'e Brillouin zone showing the high-symmetry points (defined in the text). 
    The red dashed lines show the active momentum space ``wires'' for the $L_y=6$ cylinder with $\Phi_y=0$ (PBC). These are evenly spaced in the rectangular Brillouin zone (gray rectangle). However, we often replot data onto the hexagonal Brillouin zone (black hexagon) for convenience.
    }
    \label{fig:app:real_momentum_geometry}
\end{figure*}

\section{Microscopic Model}
\label{app:microscopic_model}

We consider a simple microscopic model for \ce{tMoTe2}: a continuum model for the valence bands with screened Coulomb interactions. We now provide a self-contained description of the model.

\subsection{Lattice Geometry}

Our conventions for the TMD geometry follow~\cite{WuFengcheng2019TMD}.

We take monolayer TMD lattice vectors
\begin{align}
   &\v{R}_{1,2} = a_0 \left(\frac{1}{2}, \mp\frac{\sqrt{3}}{2}\right),&
   &\v{G}_{1,2} = \frac{2\pi}{a_0}\left(1,\mp\frac{1}{\sqrt{3}}\right),
\end{align}
where $a_0=\SI{3.52}{\angstrom}$ is the monolayer TMD lattice constant~\cite{WangChong2024FCI}. We focus on $\theta=3.89^\circ$.

Two monolayers with relative twists of $\theta/2$ and $-\theta/2$ for the top and bottom layers form a moir\'e pattern with a superlattice generated by
\begin{align}
   &\v{a}_{1} = L_M\left(\frac{\sqrt{3}}{2},\frac{1}{2}\right),&
   &\v{a}_2 = L_M(0,1),&
   &L_M = \frac{a_0}{2\sin\frac{\theta}{2}},&
\end{align}
where $L_M$ is called the moir\'e length. 
The minimal length reciprocal vectors are
\begin{align}
    &\v{b}_1 = \frac{4\pi}{\sqrt{3} L_M}\left(1,0\right),&
    &\v{b}_j = R_{(j-1)(2\pi/6)} \v{b}_1
\end{align}
where $R_{\theta}$ is the counterclockwise rotation matrix. We consider the reciprocal lattice generated by
\begin{equation}
    \v{g}_1 = \v{b}_1, \quad \v{g}_2 = \v{b}_3.
\end{equation}
The high-symmetry points of the moir\'e lattice, shown in Fig.~\ref{fig:app:real_momentum_geometry}(b), are at
\begin{align}
    \v{\kappa}_t &= \frac{1}{3}\v{g}_1 + \frac{1}{3} \v{g}_2 + \v{\gamma}\\
    \v{\kappa}_b &= -\frac{2}{3}\v{g}_1 - \frac{1}{3} \v{g}_2 + \v{\gamma}\\
    \v{\mu} &= -\frac{1}{2} \v{g}_1 + \v{\gamma}.
\end{align}

\subsection{Brillouin Zone Geometry}

Our DMRG conventions follow~\cite{Soejima2020DMRG} and~\cite{DongJunkai2023CFL}. For DMRG it is convenient to use a rectangular Brillouin zone. We use $\mathcal{B} = [-G_x/2,G_x/2] \times [-G_y/2, G_y /2]$ where $G_x = \frac{4\pi}{\sqrt{3}L_M}, G_y = \frac{2\pi}{L_M}$ so that $\v{a}_1 = 2\pi\left(  \frac{1}{G_x}, \frac{1}{2G_y} \right)$.

Instead of a Monkhorst-Pack grid, we instead take a rectangular grid of size $L_x \times L_y$, where we will take $L_x~\to~\infty$. Define grid basis vectors
\begin{align}
    \v{\delta k}_x &= \left( \frac{G_x}{L_x}, 0 \right)
    = \frac{1}{L_x} \v{g}_1,\\
    \v{\delta k}_y &= \left( 0, \frac{G_y}{L_y} \right)
    = \frac{1}{L_y}(\tfrac{1}{2}\v{g}_1+\v{g}_2).
\end{align}
We then take a $k$-grid
\begin{equation}
    \v{k}_{n_x,n_y} =
    \left(n_x + \frac{\Phi_x}{2\pi}\right) \v{\delta} \v{k}_{x} + \left(n_y + \frac{\Phi_y}{2\pi}\right) \v{\delta}\v{k}_{y}
\end{equation}
where  $0 \le \Phi_{x,y} \le 2\pi$ and 
\begin{align}
    n_x &\in [-\floor{L_x/2}, \floor{(L_x-1)/2}],\\
    n_y &\in [-\floor{L_y/2},\floor{(L_y-1)/2}]
\end{align}
For example, for $N_y = 5$, $n_y = -2,-1,0,1,2$ and for $N_y = 6$, $n_y = -3,-2,-1,0,1,2$. When $L_x \to \infty$, this gives $L_y$ evenly spaced ``wires'' in the Brillouin zone, shown in Fig.~\ref{fig:app:real_momentum_geometry}(c).

The phase $\Phi_y$ controls the boundary conditions. If $\Phi_y =0$, then one wire will go through $\v{\Gamma}$, giving periodic boundary conditions (PBC); $\Phi_y = \pi$ corresponds to anti-periodic boundary conditions (APBC). We use APBC unless otherwise stated.

To understand the real-space geometry of the whole system, we define
\begin{align}
    \v{L}_x &= N_x\left(\frac{2\pi}{G_x}, 0\right)
    = N_x (\v{a}_1 -\tfrac{1}{2} \v{a}_2),\\
    \v{L}_y &= N_y\left(0, \frac{2\pi}{G_y}\right)
    = N_y \v{a}_2.
\end{align}
which satisfy $\v{L}_a \cdot \v{\delta k}_b = 2\pi \delta_{ab}.$ 
The whole system is then a torus spanned by $\v{L}_x, \v{L}_y$, making it rectangular. We require that the unit cell torus spanned by $\v{a}_1,\v{a}_2$ tiles the whole system, which imposes the requirement
\begin{equation}
    L_x \pmod{2L_y} \equiv 0.
\end{equation}
We may then take $L_x \to \infty$, turning the torus into an infinite cylinder.

\begin{figure*}
    \centering
    \includegraphics[width=\linewidth]{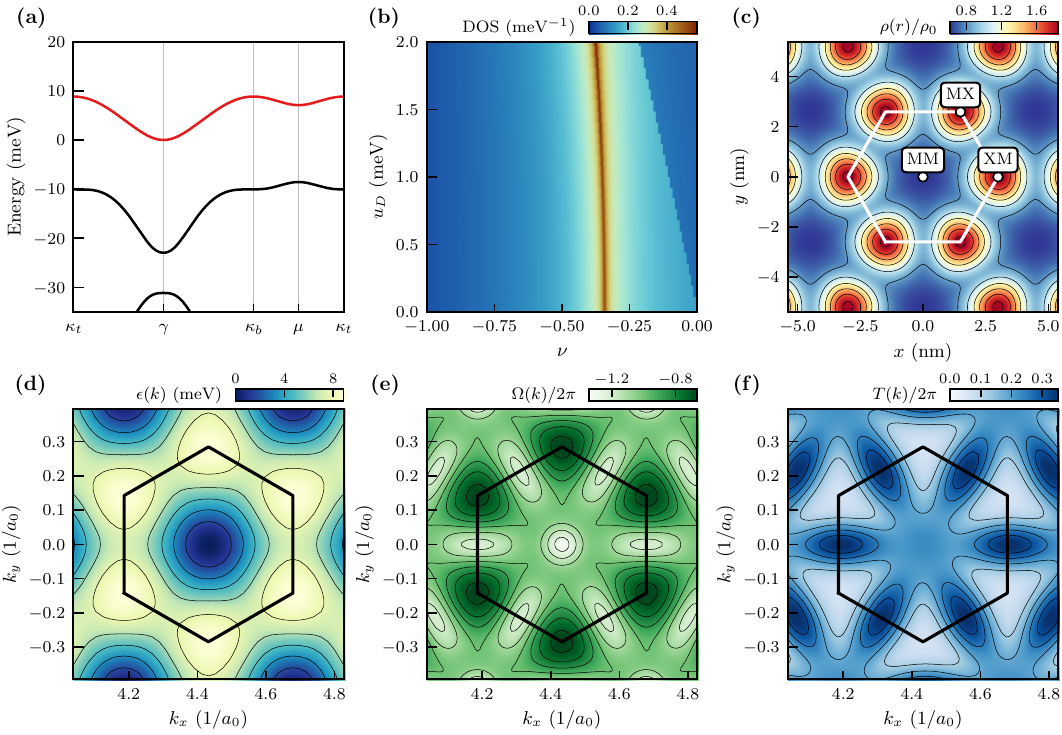}
    \caption{
    \textbf{Single particle properties of \ce{tMoTe2} at $\theta=3.89^\circ$.}
(a) Bandstructure of \ce{tMoTe2} in the $K$ valley. The zero point of energy is taken to be the minimum of the top valence band, which has bandwidth $w = \SI{8.85}{meV}$ and is separated from the second-to-top band by a bandgap of $\Delta E= \SI{8.55}{meV}$.
(b) Density of states for the top band assuming valley-polarization as a function of filling $\nu$ and displacement field $u_D$. The DOS is regulated with a finite-temperature broadening of \SI{0.003}{meV} (i.e. \SI{35}{mK}). The van Hove is at $\nu \sim -0.35$.
(c) Real space charge density of the top band, whose density is concentrated at the MX and XM Wyckoff centers. Here $\rho_{0}=-1/A_{\mathrm{uc}}$ is the average charge density at $\nu=-1$.
(d) Dispersion $\epsilon(k)$ of the top valence band, assuming valley polarization. The band minimum is at $\gamma$, which we set to $\epsilon(\gamma)=0$, with maxima at $\kappa_t$ and $\kappa_b$. The dispersion is 6-fold symmetric due to $C_{3z}$ and pseudoinversion $\mathcal{I}$.
(e) Berry curvature $\Omega(\v{k})/2\pi$, normalized so its mean value of the Brillouin zone is the Chern number $C=-1$.
(f) Trace condition violation $T(\v{k}) = \Omega(\v{k}) - \tr g(\v{k})$, whose value is small across the entire Brillouin zone.
    }
    \label{fig:app:bandstructure}
\end{figure*}

\subsection{Single-Particle Hamiltonian}
We use a standard continuum model~\cite{WuFengcheng2019TMD} for the valence bands of \ce{tMoTe2}, which we reiterate here for the reader's convenience. The model is identical to the one from previous DMRG work~\cite{DongJunkai2023CFL}. The kinetic term for the $K$-valley acts in the bottom/top layer basis ($\ell = t,b$) as
\begin{equation}
    \label{eq:app_continuum_Hamiltonian}
    \hat{h}_K(\rB) =
    \begin{bmatrix}
 h_t(\rB)+\Delta_{t}(\v{r}) + u_D & T(\rB)\\
 T^\dagger(\rB) & h_b(\rB) + \Delta_{b}(\v{r})  -u_D
\end{bmatrix}
\end{equation}
where $u_D$ is the vertical displacement field,
\begin{equation}
    h_\ell = - \frac{(\v{p} - \hbar \v{K}_\ell)^2}{2m^*}
\end{equation}
is the quadratic band maxima of the TMD valence bands,
\begin{equation}
    \Delta_{\ell}(\v{r}) = 2V \sum_{j=1,3,5} \cos(\v{b}_j \cdot \v{r} + \ell \psi)
\end{equation}
is the on-layer moir\'e potential, and
\begin{equation}
    T(\v{r}) = w (1+e^{i\v{b}_2\cdot\v{r}} +e^{i \v{b}_3 \cdot \v{r}}).
\end{equation}
We take parameters from first principles calculations of \ce{MoTe_2}~\cite{WangChong2024FCI}: $(a_0, m^*, V,\psi,w) = (\SI{3.52}{\angstrom},0.6m_e, \SI{20.8}{\milli\electronvolt}, -107.7^\circ, \SI{-23.8}{\milli\electronvolt})$ at $\theta = 3.89^\circ$. At these parameters, the top valence band for the $K$ valley has Chern number $C=+1$, and the second-to-top band has Chern number $C=-1$. Within a valley, the model has $C_{3z}$ and $C_{2y}\mathcal{T}$ symmetry, as well as moir\'e translation symmetry. 

We solve the kinetic Hamiltonian in the Bloch basis
\begin{align}
    &\hat{h}_{K} \ket{\phi_{\v{k}b}} = \epsilon_{\v{k}b} \ket{\phi_{\v{k}b}},&
    &\hat{T}_{\v{a}_j} \ket{\phi_{\v{k}b}} = e^{i\v{k}\cdot\v{a}_j} \ket{\phi_{\v{k}b}},&
\end{align}
where $b$ labels bands and $\hat{T}_{\v{a}_j}$ applies a (passive) translation by $\v{a}_j$. In practice, we use a plane-wave basis with six hexagonal shells of planewaves, as needed for convergence. Fig.~\ref{fig:app:bandstructure} shows the bandstructure, dispersion, and real-space charge density of the top band. Additionally, it computes the Berry curature and trace condition violation across the Brillouin zone finding, as in prior work~\cite{DongJunkai2023CFL}, that the band is perturbatively close to ideal at these parameters.

\subsection{Many-Body Hamiltonian}
We define electron creation operators  $\hat{c}^\dagger_{\v{k}b}$ that create the Bloch states $\ket{\psi_{\v{k}b}}$. We consider an interacting Hamiltonian
\begin{equation}
    \hat{H} = \hat{h} + \frac{1}{2A}\sum_{\v{q}} V_{\v{q}} :\hat{\rho}_{\v{q}} \hat{\rho}_{-\v{q}}:
\end{equation}
where
\begin{itemize}
    \item $:\hat{O}:$ denotes normal ordering with respect to charge neutrality (where all the valence bands are fully filled) 
    \item the kinetic term is
\begin{equation}
    \hat{h} = \sum_{\v{k} \in \mathcal{B}} \sum_{b,b'} \hat{c}^\dagger_{\v{k}b} [h(\v{k})]_{b,b'} \hat{c}_{\v{k},b};
\end{equation}
    \item in terms of the system area $A$ and relative momentum $\v{q}$ (not restricted to the Brillouin zone), the screened Coulomb interactions are
    \begin{equation}
        V_q = \frac{\tanh(q d_s)}{2\epsilon_r \epsilon_0 q}
    \end{equation}
    where $\epsilon_r$ is the relative dielectric constant and $d_s$ is both the sample-to-top-gate and sample-to-bottom-gate distance~\cite{Soejima2020DMRG}
    \item the density operators are defined as
    \begin{equation}
        \hat{\rho}_{\v{q}} 
        = \sum_{\v{k} \in \mathcal{B}}
        \hat{c}^\dagger_{\v{k},b}
        [\Lambda_{\v{q}}(\v{k})]_{bb'}
        \hat{c}_{\v{k}+\v{q},b'}
    \end{equation}
    in terms of the form factors
    \begin{equation}
        [\Lambda_{\v{q}}(\v{k})]_{bb'} = \braket{\psi_{\v{k},b}|e^{-i\v{q}\cdot\v{r}}|\psi_{\v{k}+\v{q},b'}}.
    \end{equation}
\end{itemize}
For numerics, we project this Hamiltonian into the top valence band in the $K$ valley. We have chosen conventions so that this projection is exactly the \textit{restriction} of the band sums to a single index. In terms of the dispersion $\epsilon_{\v{k}}$, form factors $\lambda_{\v{q}}(\v{k})$, and creation operators for the top band, the projected Hamiltonian is
\begin{align}
\label{eq:app_many-body_Hamiltonian}
    \hat{H}_{\text{proj}} &= \sum_{\v{k} \in \mathcal{B}} \epsilon_{\v{k}} \hat{c}_{\v{k}}^\dagger \hat{c}_{\v{k}}\\
    \nonumber
    &+ \sum_{\v{k},\v{k}' \in \mathcal{B}}
    \frac{V_q}{2A} 
    \sum_{\v{q}} :\hat{c}^\dagger_{\v{k}} \lambda_{\v{q}}(\v{k}) \hat{c}_{\v{k}+\v{q}} \hat{c}^\dagger_{\v{k}'} \lambda_{-\v{q}}(\v{k}')\hat{c}_{\v{k}'-\v{q}}:.
\end{align}

\subsection{Computational Basis}

\begin{figure}
    \centering
    \includegraphics[width=\linewidth]{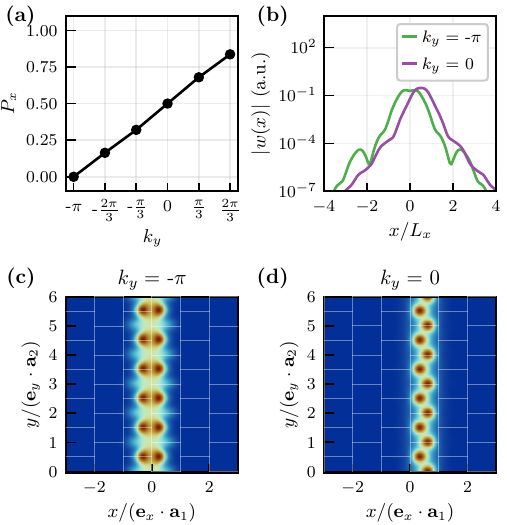}
    \caption{
    (a) Polarizations $P_x(k_y)$ on a circumference $L_y=6$ cylinder. Polarizations are defined $\pmod 1$, and $C = \int \frac{dP_x}{d k_y} dk_y = +1$ measures the Chern number.
    (b) Profile of two maximally localized wannier functions. Here $w(x) = \int dy |w(x,y)|^2$ is the integral around the cylinder.
    (c,d) Two hybrid wannier functions $|w_{k_y}(x,y)|^2$ on the $L_y=6$ cylinder. The top Wannier function, centered at $P_x(k_y = -\pi) = 0$, straddles two rings of  unit cells, while the lower one, centered at $P_x(0) = 0$, is centered in a ring of unit cells. 
    }
    \label{fig:app:wannierization}
\end{figure}

Following previous DMRG work on moir\'e systems~\cite{Kang2020DiracBraiding,Soejima2020DMRG,DongJunkai2023CFL}, we take a computational basis of hybrid Wannier functions that are maximally localized along $x$ and planewaves along $y$. In terms of the Bloch wavefunctions $\ket{\psi_{\v{k}}}$ for the top band, these are defined as
\begin{equation}
\label{eq:Wannier_states}
    \ket{w_{n,k_y}}
    = \frac{1}{\sqrt{L_x}} \sum_{k_x}  e^{-i\v{k} \cdot (n \v{a}_1)} \ket{\varphi_{\v{k}}}; \quad \ket{\varphi_{\v{k}}} =  \ket{\psi_{\v{k}}} e^{i \zeta(\v{k})}.
\end{equation}
The gauge $\zeta(\v{k})$ is chosen such that~\cite{Kang2020DiracBraiding,Soejima2020DMRG}
\begin{align}
    \hat{T}_{\v{a}_1} \ket{w_{n,k_y}} &= \ket{w_{n-1,k_y}}\\
    \hat{T}_{\v{a}_2} \ket{w_{n,k_y}} &= e^{i 2\pi k_y/G_y} \ket{w_{n,k_y}}\\
    \hat{P} e^{-i \hat{x} G_x} \hat{P} \ket{w_{n,k_y}} &= e^{i 2\pi[n+P_x(k_y)]} \ket{w_{n,k_y}}
\end{align}
where $\hat{x}$ is the position operator and $\hat{P} = \sum_{\v{k}} \ket{\psi_{k}}\bra{\psi_{\v{k}}}$ is the (single-particle) projector to the top band. Note that for wires with flux $0 \le \Phi_y \le 2\pi$, we have
\begin{equation}
    \hat{T}_{\v{a}_2}^{L_y} \ket{w_{n,k_y}} = e^{i \Phi_y} \ket{w_{n,k_y}},
\end{equation}
interpolating between periodic and anti-periodic boundary conditions. Here the polarizations are the centers of the Wannier orbitals along the x-axis:~\cite{Soejima2020DMRG}
\begin{equation}
    P_x(k_y) = \frac{G_x}{2\pi} \braket{w_{0,k_y}|\hat{x}|w_{0,k_y}} \pmod 1.
\end{equation}
(Here we have set the electric charge to be $e=1$.) The modern theory of polarization~\cite{Vanderbilt2018BerryPhases,KingSmith1993Polarization,Thouless1983Pumping} shows that the Chern number of the band is given by the winding of the polarization through the Brillouin zone as $C = \int \frac{dP_x}{d k_y} dk_y$, which here is $C=1$. Fig.~\ref{fig:app:wannierization} shows the winding the polarization, the exponential decay of the Wannier functions along $x$, and their real-space profiles.

At the many-body level we define Wannier creation operators $\hat{c}^\dag_{n,k_y}$ via (note minus signs)
\begin{equation}
\label{eq:computational_basis}
    \hat{c}^\dagger_{\v{k}} = \frac{1}{\sqrt{L_x}} \sum_{k_x}  e^{+i\v{k} \cdot (n \v{a}_1)} e^{-i\zeta(\v{k})} \hat{c}_{n,k_y}^\dagger.
\end{equation}
In the main text and below we write the computational basis as
\begin{equation}
    \hat{c}_{x,k_y}
\end{equation}
where $x = x_n = \v{e}_x \cdot (\v{a}_1 n)$ is the $x$-position of the center of the $n$th ring. Note that the exponential factor $e^{+i\v{k} \cdot (n \v{a}_1)} \neq e^{i k_x x}$, but instead has an extra phase. The many-body Hamiltonian Eq.~\eqref{eq:app_many-body_Hamiltonian} can be readily expressed in this basis. This puts the \ce{tMoTe2} Hamiltonian onto a cylinder geometry with a circumference of $L_y$ unit cells that is infinite along $x$ but composed of maximally-localized hybrid Wannier functions. This choice ensures that the kinetic hoppings decay quickly along $x$. However, the quasi-long range screened Coulomb interactions still decay slowly, requiring special techniques to perform DMRG in practice.

\subsection{MPO Compression}
To perform infinite DMRG, one must express the Hamiltonian as an infinite matrix product operator (MPO), which for fermions is a tensor network of size $2\times 2 \times D \times D$ where $D$ is the virtual bond dimension. Naive constructions of MPOs truncate the interaction range in the $\hat{d}_{n,k_y}^\dagger$ basis at some cutoff length $M$. Not only does this cutoff cause errors, potentially biasing the Hamiltonian, but it requires huge bond dimensions of $D \sim 2000$ for only $M=2$ at $L_y=6$~\cite{Soejima2020DMRG}. To avoid these issues, we apply the technique of iMPO compression developed by some of us~\cite{Parker2020MPO,Soejima2020DMRG}, which faithfully represents the full Hamiltonian including its quasi-long range interactions. In practice we check that each matrix element is represented with error $\epsilon < \SI{1e-3}{meV}$ out to separations of 200 sites along the 1d chain. The resulting iMPOs have bond dimension $D \sim 120$ for $L_y=6$, and grow modestly with $L_y$.

\section{Practical Application of The Scaling Theory of iMPS}
\label{app:MPS_scaling_theory}

\begin{figure*}
    \centering
    \includegraphics[width=\linewidth]{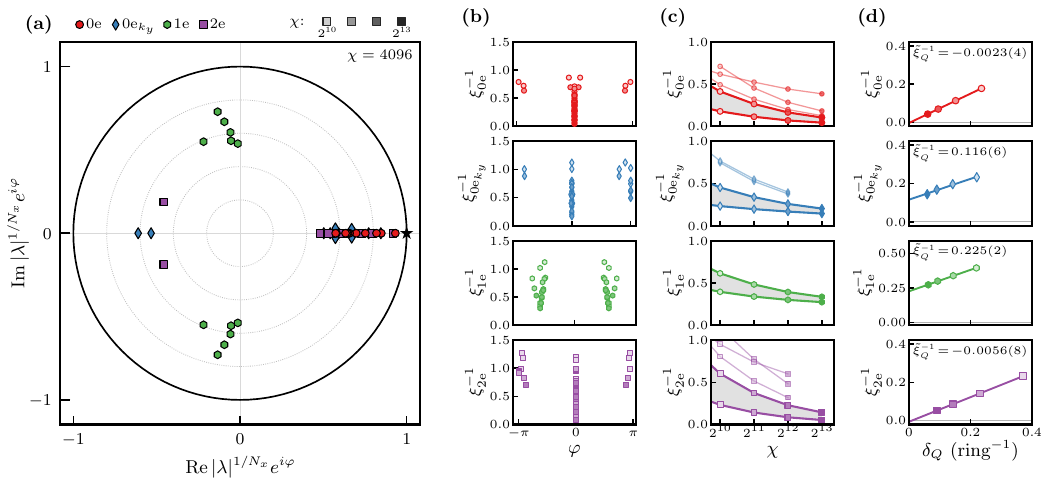}
    \caption{
    \textbf{Correlation length extrapolation.}
(a). The leading transfer matrix eigenvalues in the most important sectors at $\chi=4096$ for the parameters in Fig.~\ref{fig:2_SC_evidence}: $\nu=-0.73$, $u_D=0$, $\epsilon_r=8$, $d_s=\SI{1.5}{nm}$. Different sectors of the transfer matrix are shown by color. One can clearly see different branches going towards the outside of the circle ($\n{\lambda}=1$) at different phases $\phi$.
(b). The inverse correlation lengths $\xi_{Q}^{-1} = \epsilon_Q$ in each sector as a function of bond dimension (saturation of each color).
(c) Inverse correlation lengths versus bond dimension. Lines connect corresponding correlation lengths at different bond dimension. The gray shaded region shows the difference $\delta_Q$ between the leading and subleading correlation lengths from a particular branch.
(d) Extrapolated correlation lengths using the scaling variable $\delta_Q$ using linear fits. See Fig.~\ref{fig:2_SC_evidence} of the main text.
    }
    \label{fig:app:transfer_matrix_extrapolation}
\end{figure*}

This Appendix briefly reviews the scaling theory of matrix product states and how it may be applied in practice to extrapolate correlation lengths to bond dimension $\chi \to \infty$ in a principled manner. The essential idea is that at finite $\chi$ all MPS correlation functions decay exponentially $C_{\chi}(x) \sim e^{-x/\xi}$ --- making it impossible to encode a power law prefactor $C_{\eta}(x) \sim x^{-\eta} e^{-x/\xi}$ exactly, including the case $\xi=\infty$. However, the limit 
\begin{equation}
    C_{\chi}(x) \xrightarrow{\chi\to\infty} C_{\eta}(x) 
\end{equation}
is well-controlled \textit{in the right variable} $\delta(\chi)$. The classic references on this topic, from which we draw extensively below,  are~\cite{Rams2018Extrapolation,Vanhecke2019Scaling,Schneider2025Scaling,Hodgkiss2026Scaling}. 

Suppose we have an injective iMPS of bond dimension $\chi$ with transfer matrix
\begin{equation}
    \mathsf{T} = \sum_{j=1}^\chi \pket{R_j} \lambda_j \pbra{L_j}
\end{equation}
with $\pbraket{L_i|R_j} = \delta_{ij}$ and $1 = \lambda_0 > \n{\lambda_1} \ge \n{\lambda_2} \ge \cdots $. Here and below rounded kets $\pket{v}$ denote vectors in the doubled virtual space that the transfer matrix acts on. Fig.~\ref{fig:app:transfer_matrix_extrapolation}(a) shows an example transfer matrix spectrum.

In terms of this, the \textit{connected} correlation function 
\begin{equation}
    C_{O}(x) = \braket{\Psi|\hat{O}(x)^\dagger \hat{O}(0)|\Psi}
    - \braket{\Psi|\hat{O}(x)^\dagger |\Psi}
    \braket{\Psi|\hat{O}(0)|\Psi}
\end{equation}
takes the form
\begin{equation}
    \label{eq:sum_decaying_exponentials}
    C_{\mathcal{O}}(x) = \sum_{j>0} w_j \lambda_j^x,
\end{equation}
where $w_j$ are form factors determined by $\hat{O}$, $\pket{R_j}$ and $\pket{L_j}$. We decompose the eigenvalues into their magnitude and phase as
\begin{equation}
    \lambda_j = e^{-(\epsilon_j+i\phi_j)a} = e^{-i a \phi_j} e^{-a/\xi_j}; \quad
    \xi_j^{-1} = \epsilon_j
\end{equation}
where $a$ is the lattice spacing, $\epsilon_j \ge 0$, and $\xi_j = 1/\epsilon_j \ge 0$ is called the correlation length. Such a correlation function decays exponentially:
 $C_{O}(x) \xrightarrow{x \gg 1} w_1 \lambda_1^x = w_1 e^{-i\phi_1} e^{-x/\xi_1}$. We note the phase $\phi$ contains a wealth of physical information. For instance, for $(1e,k_y)$, it is related to the $k_x$ values where the $k_y$ wire intersects the Brilloin zone.

Correlation function often takes the form $C_{\eta}(x) \sim e^{-x/\xi} x^{-\eta}$, which cannot be represented as any finite sum of exponentials, and is therefore beyond an exact MPS representation at finite $\chi$. However, they can be represented by a continuous integral over decaying exponentials. Concretely, \begin{equation}
\label{eq:power_law_decay}
    C_{\eta}(x) = \frac{e^{-x/\xi}}{x^\eta} 
        = \int_0^{e^{-\xi}} d\lambda \; \frac{(-\ln \lambda)^{\eta-1}}{\Gamma(\eta) \lambda} \lambda^x,
\end{equation}
which follows from the definition of the $\Gamma$ function. As $\chi \to \infty$, the points $\lambda_j$ must therefore \textit{accumulate} into lines in the unit disk. We refer to these lines as ``branches'', as they produce the branch cuts in the Stieltjes transform representation of power laws like Eq.~\eqref{eq:power_law_decay}.

To quantify this, it is standard to define scaling variables with units of inverse length:
\begin{equation}
    \delta_{ij}(\chi) = \epsilon_i(\chi) - \epsilon_j(\chi),
\end{equation}
for particular $\lambda_i$ and $\lambda_j$. As $\chi \to \infty$, $\delta_{ij}(\chi) \to 0$. This can be thought of as the small length scale at which the power laws is gives in to the exponential decay~\cite{Schneider2025Scaling}. While appealing, this leaves open an obvious practical question: \textit{which} $\lambda_i$ and $\lambda_j$ should be used?

To illustrate the difficulty, suppose for an operator $\hat{O}$ with charge $Q$, the true ground state correlation takes the form:
\begin{align}
    C_O(x) &= P_1 e^{i \phi_1 x} x^{-\eta_1} + P_2 e^{i \phi_2 x} x^{-\eta_2} +  \cdots\\
    &+ E_1 e^{i \phi_3 x } x^{-\eta_3} e^{-x / \xi_3} + E_2 e^{i \varphi_4 x } e^{-\eta_4} e^{-x / \xi_4} + \cdots,
\end{align}
which has both multiple power laws (potentially with different phases $\phi_j$) and multiple exponential terms. Each of these functions will give rise to a different branch in the unit disk on which eigenvalues will accumulate.
Our practical prescription is then as follows: isolate the eigenvalues belonging to each branch in the correlation function via a careful inspection of the transfer matrix eigenvalue spectrum and take $\delta_{ij}$ corresponding to separation between the largest ones. 

We note several practical issues. First, $\lambda_i$ and $\lambda_j$ should necessarily be from the same symmetry sector. This is easy to enforce when the symmetry is conserved, but additional non-conserved symmetries (perhaps non-Abelian ones) can also act on the transfer matrix spectrum. Second, the phase of $\lambda_i$ is often helpful in identifying a branch, but multiple branches can have the same phase. Third, if there are ``bystander'' exponential correlations like $E_1 \lambda_e^x =  E_1 e^{i \varphi_1 x } e^{-x / \xi_1}$ with long correlation lengths, then the power law eigenvalues $\lambda_j$ will cross $\lambda_e$ as $\chi \to \infty$. The index $j$ for the scaling variable $\delta_{ij}$ changes at this point, and we must use continuity to assign $j$ correctly.
Fourth, even within the eigenvalues $\lambda_j$ for a particular power law branch, there may be finite-$\chi$ eigenvalue crossings. For instance, with the fourth and fifth eigenvalues swapping places between $\chi=2048$ and $\chi=4096$. An avoided crossing will create deviations from the scaling behavior, while an unavoided crossing will create a labeling issue. Only once such issues are accounted for can a practical identification of $\delta_{ij}(\chi)$ be made for each $\chi$ and used for extrapolating to $\chi\to\infty$. This is shown in Fig.~\ref{fig:app:transfer_matrix_extrapolation}(c) where, for instance, several eigenvalue crossings are visible in the higher correlation length values. 

The fits used in practice are shown in Fig.~\ref{fig:app:transfer_matrix_extrapolation}(d). Other points are fit using a similar procedure. As $\xi^{-1} \ge 0$, the slightly negative extrapolated values like $-0.0023$ are an error incurred by the extrapolation procedure.

\section{Proxy Pair Wavefunction}
\label{app:proxy_pair_wavefunction}

In this section we define the ``proxy pair wavefunction'' defined in the main text and argue that it reflects the pairing symmetry of the superconductor in the 2d limit. This section draws from, but also expands upon, previous work from some of us on this topic~\cite{Sahay2024Superconductivity,Divic2025AnyonSC,WangTaige2025ChiralSC}.

Consider the 2d Cooper pair operator
\begin{equation}
    \hat{\Delta}(\v{r};\delta\v{r}) = \hat{c}_{\v{r}+\delta\v{r}/2}\hat{c}_{\v{r}-\delta\v{r}/2},
\end{equation}
and let $\hat{\Delta}(\v{r};\v{k})$ be the Fourier transform in its relative coordinate. The ground state wavefunction of a 2d superconductor spontaneously breaks electromagnetic symmetry $U(1)_Q$, whose order parameter is the pair wavefunction
\begin{equation}
     \Delta(\v{k}) = \braket{\hat{\Delta}(\v{r}; \v{k})}_{2d},
\end{equation}
whose value is $\v{r}$-independent for pairing at zero momentum, which we assume throughout. The transformation of the pair wavefunction $\Delta(\v{k})$ under the space group, particularly rotations, determine the superconducting pairing channel. Within DMRG, however, we only have access to cylinder wavefunctions $\Psi(L_y,\chi)$ that preserve $U(1)_Q$. How may $\Delta(\v{k})$ be probed from that data?

\begin{figure*}
    \centering
    \includegraphics[width=\linewidth]{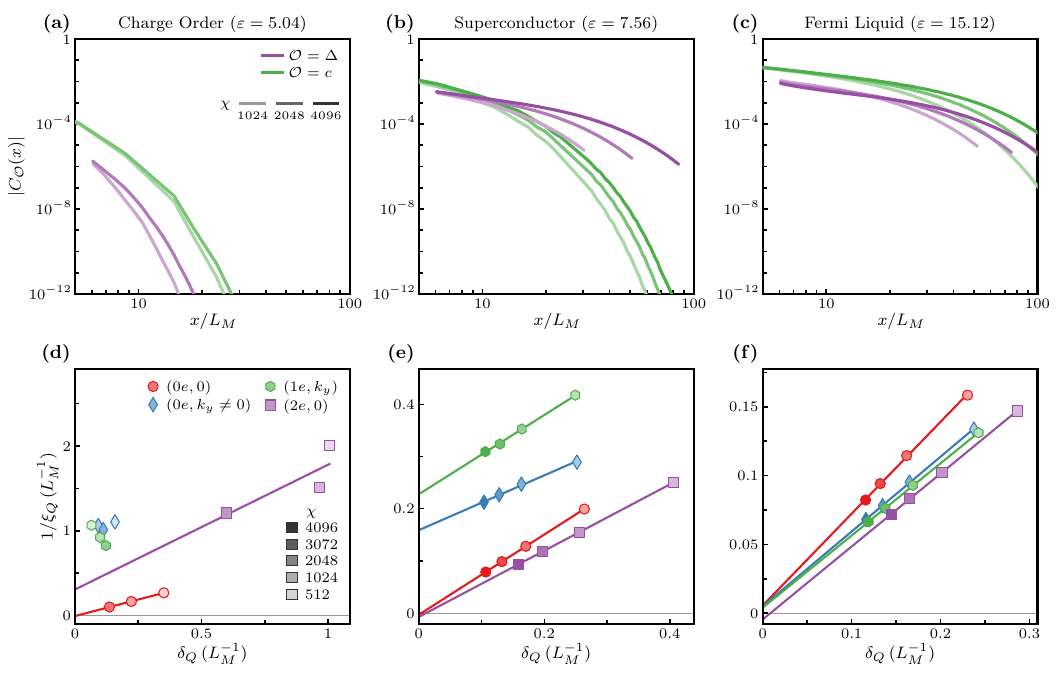}
    \caption{
    Correlations and correlation lengths of the three representative parameter points specified in Table.~\ref{tb:representative_parameter_points}.
    (a,b,c) Electron-electron and Cooper-Cooper correlations in each state for several bond dimensions. The electron-electron correlations are oscillatory, so an upper-envelope is used by subsampling points. For the Cooper-Cooper correlations, the largest singular value $\sigma_0(x)$, defined in Eq.~\eqref{eq:Cooper_correlation_singular_value}, is plotted.
    (d,e,f). Extrapolation of correlation lengths with the scaling variable $\delta_Q$ in four relevant channels. Extrapolation is carried out as in App.~\ref{app:MPS_scaling_theory}. The identity of the phases is discussed in App.~\ref{app:phase_identification}. To represent the correct phase, (a,d) present the RTS state and the others present the MTS states. We note that the state in (a,d) is not representative of the thermodynamic limit.
    }
    \label{fig:app:phase_identification_extrapolation}
\end{figure*}

We briefly recapitulate the argument from~\cite{Sahay2024Superconductivity}, eliding details to streamline the discussion.
In the far-field regime $\n{\v{r}} \gg |\delta \v{r}|$, cluster decomposition (indirectly) implies
\begin{equation}
\label{eq:SC_long_range_order}
    \underbrace{\braket{\hat{\Delta}^\dagger(\v{r};0) \hat{\Delta}(0;\delta \v{r})}}_{D(\v{r};\delta \v{r})}
    = \underbrace{\braket{\hat{\Delta}^\dagger(\v{r};0)}}_{A(\v{r})} \underbrace{\braket{\hat{\Delta}(0;\delta \v{r})}}_{\Delta(\v{r})},
\end{equation}
which is an expression of long-range order. (Here and below, $=$ is equality in the far-field regime) If we fix a particular $\v{r}$ in the far-field regime, we may therefore compute the pair wavefunction up to an overall constant as
\begin{equation}
    \Delta(0;\delta \v{r}) = \frac{D(\v{r};\delta \v{r})}{A(\v{r})}.
\end{equation}
If we now go to finite-but-large circumference $L_y$, long range order becomes quasi-long range order
\begin{equation}
    \braket{\hat{\Delta}^\dagger(\v{r};0) \hat{\Delta}(0;\delta \v{r})}_{L_y}
    = \frac{D(x,y;\delta \v{r}|L_y)}{\n{x}^{\eta(L_y)}} + \cdots.
\end{equation}
For sufficiently large $L_y$, one expects the limit to be smooth, with
\begin{equation}
    D(x,y;\delta \v{r}|L_y) = D(\v{r};\delta \v{r}) + O(1/L_y)
\end{equation}
in the far-field regime $x \gg \delta x$ and, of course, $0 < y < L_y$. Finally, at finite but large bond dimension, one expects
\begin{align}
    &\braket{\hat{\Delta}^\dagger(\v{r};0) \hat{\Delta}(0;\delta \v{r})}_{L_y,\chi}\\
    &\qquad = \begin{cases}
        \displaystyle\frac{D(x,y;\delta \v{r}|L_y)}{\n{x}^{\eta(L_y)}} & \delta x \ll x < 1/\delta_{2e}\\
        A(y, \delta \v{r}) e^{-x/\xi_{2e}}  & x \gg 1/\delta_{2e}
    \end{cases}
\end{align}
where $\delta_{2e}(\chi)$ is the scaling variable from Sec.~\ref{app:MPS_scaling_theory} that approximately sets the crossover from power law to exponential. That is, in the intermediate regime of far field by less than $1/\delta_{2e}$, the finite bond dimension correlation function matches the infinite bond dimension correlation function. We have thus argued
\begin{align}
    \Delta(0;\delta \v{r}) &= \frac{D(\v{r};\delta \v{r})}{A(\v{r})} & \text{(far field) }\\
    &\approx \frac{D(\v{r};\delta \v{r}|L_y)}{\n{x}^{\eta(L_y)} A(\v{r})} & \text{ (far field)}\\
    &\approx \frac{D(\v{r};\delta \v{r}|L_y,\chi)}{\n{x}^{\eta(L_y)} A(\v{r})} & \text{ (power law regime).}
\end{align}
As $\eta(L_y)$ and $D(\v{r};\delta \v{r}|L_y,\chi)$ are directly computable from DMRG data, one may use them to approximate $\Delta(0;\delta \v{r})$ and, by Fourier transforming, $\Delta(\v{k})$. We emphasize the above is an argument where error bounds are not made precise, though each step is physically reasonable.

We now move beyond previous references. By Eq.~\eqref{eq:sum_decaying_exponentials}, the four-point function takes the form
\begin{equation}
\frac{D(\v{r};\delta \v{r}|L_y,\chi)}{\n{x}^{\eta(L_y)}} \approx \sum_{j} w_j(y, \delta \v{r}) \lambda_j^x
\end{equation}
where $\lambda_j$ are the transfer matrix eigenvalues in the 2e sector, which depend on $L_y$ and $\chi$, but are independent of $\v{r} = (x,y)$ and $\delta \v{r}$. In the power-law regime, the sum is dominated by the eigenvalues from the branch of the transfer matrix spectrum responsible for the power law:
\begin{equation}
    \sum_{j} w_j(y, \delta \v{r}) \lambda_j^x \approx \sum_{j \in \text{2e branch}} w_j(y, \delta \v{r}) \lambda_j^x.
\end{equation}
One property is immediately obvious: as these terms limit to a single power law, each of the $w_j(y, \delta \v{r})$'s must transform under the same irrep of the space group. A single $w_j$, in particular $w_1$, is therefore sufficient to determine the pairing symmetry.

We therefore define the ``proxy pair wavefunction'' to be
\begin{equation}
    \widetilde{\Delta}(\delta \v{r}) = w_1(0; \delta \v{r}),
\end{equation}
which is Fourier transformed in the relative coordinate to yield $\widetilde{\Delta}(\v{k})$ in the main text. We have checked that using $w_2$ results in a small quantitative error, confirming the robustness of this proxy. Further, $\widetilde{\Delta}(\delta \v{r})$ is significantly cheaper to compute than $D(\v{r};\delta \v{r}|L_y,\chi) \n{x}^{-\eta(L_y)}$, which was used in previous work~\cite{Sahay2024Superconductivity}.

\begin{figure*}
    \centering
    \includegraphics[width=\linewidth]{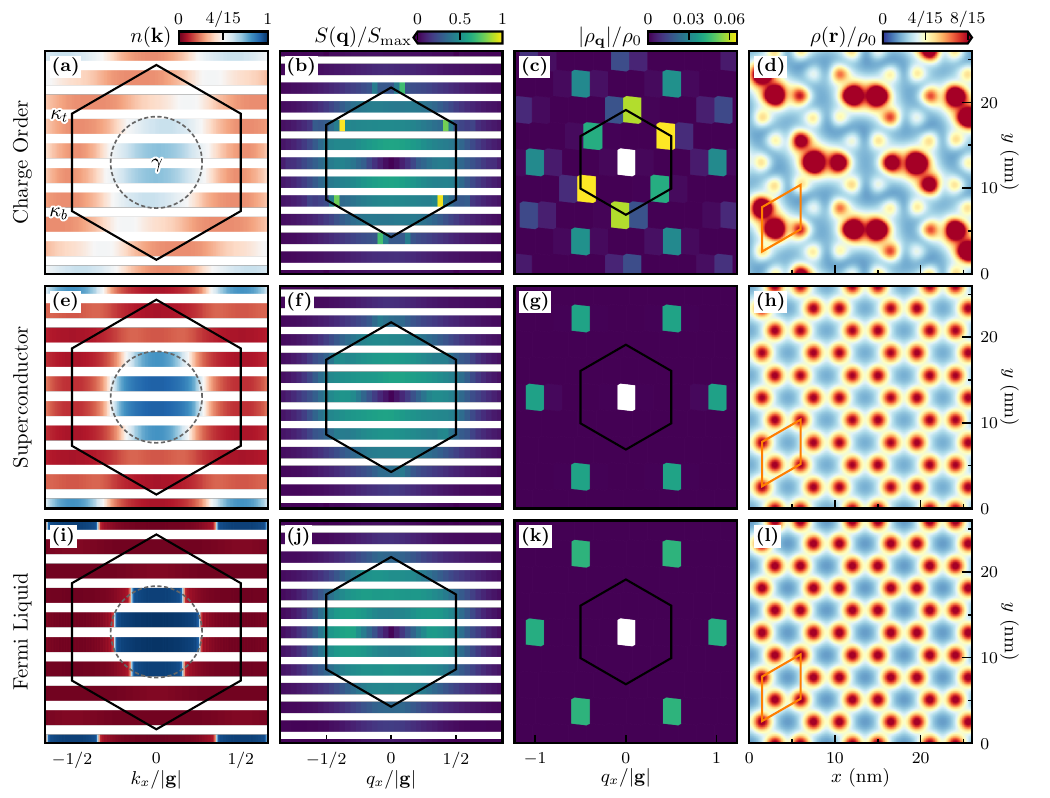}
    \caption{
    \textbf{Observables in each phase.} Selected observables at the representative points in each phase specified in Table~\ref{tb:representative_parameter_points}. Details in text.
(a,e,i) The occupations $n(\v{k})$ over the Brillouin zone at $\chi=2048$ (RTS) in (a) and $\chi=4096$ (MTS) in (e,i).
(b,f,j) Structure factor $S(\v{q})$ as a function of unrestricted momentum $\v{q}$ at $\chi=2048$ (RTS) in (b) and $\chi=4096$ (MTS) in (f,j).
(c,g,k) Fourier transform of the real-space charge density $\rho(\v{q})$ at $\chi=2048$ (RTS). Only $\v{q}=\v{G}$ in the reciprocal lattice set by translation symmetry on the cylinder are allowed, a set that does not include $\v{\kappa}_{t,b}$.
(d,h,l) Real-space charge density $\rho(\v{r})$ at $\chi=2048$ (RTS). The orange parallelogram shows the moir\'e unit cell.
We note that the state in (a-d) is not representative of the thermodynamic limit.
    }
    \label{fig:app:phase_identification}
\end{figure*}

\section{DMRG Details \& Phase Identification}
\label{app:phase_identification}
In this section we describe the details of how each phase is identified. We first describe the DMRG computations that were performed, and the observables used for their analysis.

\subsection{DMRG Computations}

DMRG computations were performed at a variety of fillings, displacement fields, screening lengths, and dielectric constants, as shown in the figures. To permit a filling $\nu=p/q$, the iDMRG unit cell must obey $q | N_x N_y$ (i.e. $Nx Ny \pmod q = 0$), necessitating different choices of $N_x$ for different fillings. Some phases also required special constraints. For example, Laughlin FCIs require $3 | N_x$ to permit all three root states. In practice, our representative points use $N_x=5$ and $N_y=6$, giving 30 total sites within the iDMRG unit cell, which supports filling $\nu=-22/30 = -11/15 \approx -0.73$.

DMRG computations were performed at each parameter point as various bond dimension $\chi$, reaching at least $\chi=2048$ and sometimes as large as $\chi=12288$. (Used for example at largest $d_s$.) For most points, two sets of DMRG computations were performed: those conserving the maximal translation symmetry, and those preserving only a subgroup of translation symmetry so as to allow $\sqrt{3}\times\sqrt{3}$ order to develop. For instance, at $N_y=6$, these would have translation by one unit cell around the cylinder, conserving momentum $6$-fold momentum, versus translation by three unit cells around the cylinder, conserving $2$-fold momentum. The Hilbert space with reduced translation symmetry (RTS) is a strict superset of the maximal translation symmetric (MTS) space, so the variational energy is expected to be strictly smaller from DMRG in this space at the same bond dimension. This is indeed observed in practice. However, the reduced translation symmetry makes the computations substantially more expensive, even at the same bond dimension. We therefore use the RTS states to analyze the formation of charge order, but perform the analysis of non-charge ordered states with MTS states.

\begin{figure*}
    \centering
    \includegraphics[width=\linewidth]{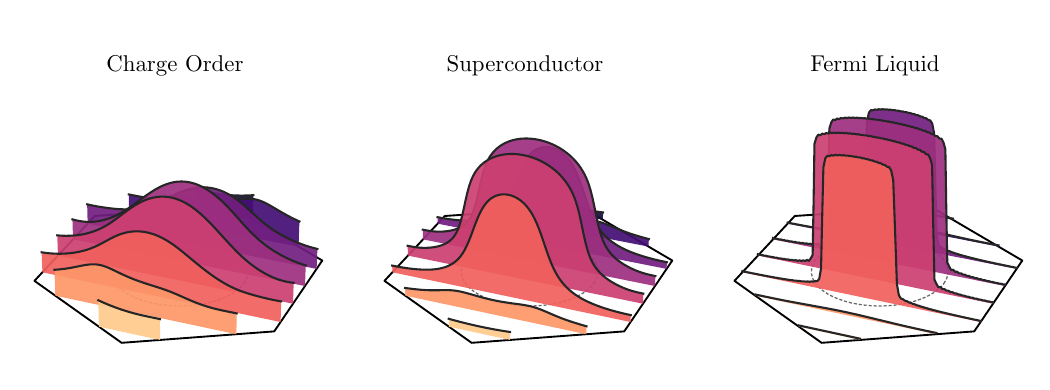}
    \caption{Three-dimensional representation of the DMRG occupations $n(\v{k})$ in the Brillouin zone. Data and parameters match Fig.~\ref{fig:app:phase_identification}(a,e,i). We note that the state in (a,d) is not representative of the thermodynamic limit.}
    \label{fig:app:occuaptions_artistic}
\end{figure*}

\subsection{DMRG Observables}
For each parameter and each RTS or MTS, we monitor the convergence of the state across a variety of difference convergence parameters and physical observables to ensure it is well-converged and meaningful. Phase assignment is then made computing a basket of physical observables, which are examined manually to ensure they are numerically and physically reasonable for each point. The most important observables are: the correlation lengths in the $0e$, $1e$, and $2e$ sectors, whose extrapolation was described in App.~\ref{app:MPS_scaling_theory} above; the momentum-space occupation $n(\v{k})$; the real-space charge density $\rho(\v{r})$; its Fourier transform $\rho(\v{q})$; the structure factor $S(\v{q})$; and the Cooper-Cooper correlation function $D$. 

Most of these are simple. For instance,
\begin{equation}
    n(\v{k}) = \braket{\hat{c}^\dagger_{\v{k}} \hat{c}_{\v{k}}},
\end{equation}
which is easily expressed in terms of the computational basis using Eq.~\eqref{eq:computational_basis}. Similarly,
\begin{equation}
    \rho(\v{r}) = \sum_{\v{G}} \rho_{\v{G}} e^{-i\v{G}\cdot\v{r}}; \quad \rho_{\v{G}} = \sum_{\v{k}} n(\v{k}) \Lambda_{\v{G}}(\v{k}),
\end{equation}
where, due to Poisson summation, the only allowed momentum points are at $\v{q}=\v{G}$ in the reciprocal lattice determined by the translation symmetry of the state. The structure factor is more complex, but was described in detail in Ref.~\cite{DongJunkai2023CFL}. 

To define the pair correlation function $D$, we first define a Cooper operator
\begin{equation}
    \hat{\Delta}(X,x,k_y) = \hat{c}_{X+\ceil{\tfrac{x}{2}},k_y} \hat{c}_{X-\floor{\tfrac{x}{2}},-k_y}
\end{equation}
where $X$ is the center of mass position, $x$ is the relative position between the pairs, and $k_y$ is their momentum. (We specialize to pairing at zero total momentum.) In terms of these, we define the Cooper-Cooper correlation function 
\begin{align}
    &D(X,x,k_y,x',k_y')
    = \braket{\hat{\Delta}(0,x,k_y)^\dagger \hat{\Delta}(X,x',k_y)}\\
    &= \Braket{
    \hat{c}^\dagger_{\ceil{\tfrac{x}{2}},k_y} \hat{c}^\dagger_{-\floor{\tfrac{x}{2}},-k_y}
    \hat{c}_{X+\ceil{\tfrac{x'}{2}},k_y'} \hat{c}_{X-\floor{\tfrac{x'}{2}},-k_y'}
    },
\end{align}
which is defined for $X > (x+x')/2$ so the annhilation operator is strictly to the right of the creation operator. (Outside this regime, various minus signs are required, whose details we elide.)

In a superconductor (or metal), one expects that at least some components of this correlation function have power-law behavior in the large bond dimension limit: $D(X) \sim X^{-\eta_{SC}}$. However, individual components do not necessarily behave in such a nice manner. For instance, they can have additional oscillations $e^{i\zeta X} X^{-\eta_{SC}}$, or can be a mix of exponentially decaying and power law decaying correlations in different $(x,k_y)$ channels. Physically, however, one expects there to be a ''Cooper operator'' which is some particular superposition $\hat{\Delta}(X) = \sum_{x,k_y} u_{x,k_y} \hat{\Delta}(X,x,k_y)$ that is a low-energy excitation in the superconductor whose structure should dominate the long-range correlations.

Based on such intuition, we define superindices $I = (x,k_y)$ and $I' = (x',k_y')$, whereupon the Cooper-Cooper correlation function becomes a matrix $[\mathsf{D}(X)]_{I,I'} = D(X,x,k_y, x',k_y')$ at each $X$. We then perform a singular value decomposition
\begin{equation}
    \mathsf{D}(X) = \mathsf{U}(X) \mathsf{S}(X) \mathsf{V}^\dagger(X), \quad \mathsf{S}(X) = \operatorname{diag} \st{\sigma_j(X)},
\end{equation}
where $\sigma_j(X)$ are the singular values of $\mathsf{D}(X)$. In practice we therefore plot
\begin{equation}
\label{eq:Cooper_correlation_singular_value}
    \sigma_0(X) = \max_{j} \sigma_j(X),
\end{equation}
the largest singular value, as a function of the center of mass separation. The main text plots labelled $\braket{\hat{\Delta}(x)\hat{\Delta}(0)}$ indeed plot $\sigma_0(x)$. We find that the corresponding singular vector $\v{u}_0(X)$ converges quite quickly with $X$ in practice, in line with the above intuition.

\begin{table}[t]
    \centering
    \begin{tabular}{ccccc}
    \toprule
    Point & $\epsilon_r$ & $d_s$ (nm) & $U$ (meV) & Phase Identification\\
    \colrule    
A     & 5.04 & 1.4 & 39 & Charge Order\\
B     & 7.56 & 1.4 & 26 & Superconductor\\
C     & 15.12 & 1.4 & 13 & Fermi Liquid\\
\bottomrule
    \end{tabular}
    \caption{Representative parameter points studied in the phase identification section. All points are examined at filling $\nu=-0.73$, displacement field $u_D=0$, and $\theta=3.89^\circ$. Here $U = \n{n}V_{q=0}(d_s,\epsilon_r)/2$ is a measure of the interaction strength.}
    \label{tb:representative_parameter_points}
\end{table}

\subsection{Phase Identification}

We focus on three representative parameter points whose only difference is the relative dielectric constant $\epsilon_r$, while screening length $d_s$, filling, etc are held fixed. The parameters of the points are described in Table~\ref{tb:representative_parameter_points}.

Fig.~\ref{fig:app:phase_identification_extrapolation} shows the key correlation functions and the extrapolation the correlation lengths in various sectors across the three parameter points. The Fermi liquid is clearly identified: its 1e and 2e correlation functions have clear power-law behavior, and the inverse correlation lengths $\xi_{0e}^{-1},\xi_{1e}^{-1},\xi_{2e}^{-1}$ all vanish upon extrapolation, as expected. The superconducting data shown here recapitulates the phenomenology of Fig.~\ref{fig:2_SC_evidence}.

Finally, the point at $\epsilon=5.04$ shows much shorter-range correlation in 1e and 2e, and only $\xi_{0e}^{-1}$ vanishes for RTS state. We interpret the state as forming a charge order that gaps out the $1e$ and $2e$ charge excitations. We note that the unit cell size here does not allow $\sqrt{3}\times\sqrt{3}$ order. Indeed, the characteristics of this state are consistent with a gapped one dimensional phase with a unit-cell-size dependent density wave, whose thermodynamic limit is ill-defined. This has the effect of gapping out charge excitations, potentially spuriously. Future work will examine states where $\sqrt{3}\times\sqrt{3}$ order is possible.

Fig.~\ref{fig:app:phase_identification} shows key observables in each state. For the Fermi liquid, the occupations change sharply from $n(k) = 0$ to $n(k) =1$ across a circular Fermi surface centered at $\gamma$. The structure factor look roughly as expected for a Fermi liquid, and the charge density is peaked strongly at the MX and XM centers shown in Fig.~\ref{fig:app:bandstructure}(c), with no additional translation breaking. 

The superconductor looks qualitatively similar to the metal in these observables, which is to be expected in a quasi-1d system. The occupations $n(\v{k})$ show the clearest distinction, with a relatively smooth crossover from empty to occupied near the Fermi surface. Here our phase identification therefore depends more strongly on the extrapolated correlation lengths.

The charge-ordered state, by contrast, is immediately apparent from $\rho(\v{r})$. Due to the $(N_x,N_y)=(5,6)$ iDMRG unit cell, the $\v{\kappa}_{t,b}$ points are not accessible momenta. However, $\rho(\v{q})$ is strongly peaked in the momenta closest to $\v{\kappa}$, signaling charge order. As the density is incommensurate with a $\sqrt{3}\times\sqrt{3}$ charge order, the peak in $\rho(\v{q})$ is slightly offset from $\v{\kappa}$ to realize a charge order commensurate with $(N_x,N_y)$. At accessible bond dimensions, correlations in the $Q=1e$ sector remain gapped. (We note that going beyond $\chi=2048$ incurs significant expense when translations are broken to permit charge order.) The spatial patterns look somewhat similar to the fidget spinner patterns in Fig.~\ref{fig:5_FCI-CDW}, except that many spinners now have a ``dimer''-like center with two peaks or a ``water molecule'' like center with three peaks. The strange and inconsistent features of this state will be topic of future work. 

Finally, Fig.~\ref{fig:app:occuaptions_artistic} provides a three-dimensional plot of the occupations across the Brillouin zone in each phase.

\section{Plane Wave Excitations}

To obtain the charge gap for 1e and 2e excitations, we computed the excitation spectrum $\omega_\pm(\boldsymbol{k})$ using the plane-wave quasiparticle ansatz MPS~\cite{Haegeman2012Excitations,Vanderstraeten2019Tangent}, with $\pm$ denoting the addition and removal of respective particles. The momentum of the excitations $\boldsymbol{k}=(k_x,k_y)$ is determined by the MPS charge number $k_y(n_y)$ and phase $k_x = (\theta+j)/N_{uc} \mod 1, j=0,1,2, N_{uc}-1$, where $\theta$ is the phase difference between sites of the plane wave superposition ansatz, and $N_{uc}$ is the size of the enlarged unit cell. The charge gap is consequently defined as $\Delta = \min_{\boldsymbol{k}}\omega_+(\boldsymbol{k})- \max_{\boldsymbol{k}}\omega_-(\boldsymbol{k})$.

\begin{figure}
    \centering
    \includegraphics[width=\linewidth]{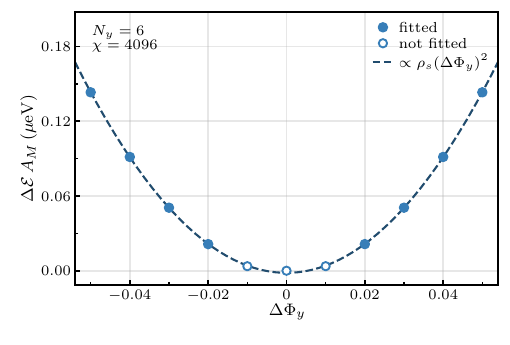}
    \caption{
    Ground-state energy change per moir'e unit cell $A_M\Delta\mathcal E$ as a function of boundary-condition twist $\delta\Phi_y$. The quadratic fit to Eq.~\eqref{eq:SC_stiffness} excludes the three central points (empty circles) and gives $\rho_s=\SI{30.5}{\micro\electronvolt}$. Parameters are $N_y=6$, $\chi=4096$, $\nu=-0.733$, $u_D=0$, $d_s=\SI{1.4}{nm}$, and $\epsilon_r=7.56$. 
    }
    \label{fig:app:stiffness_fitting}
\end{figure}

\begin{figure}
    \centering
    \includegraphics{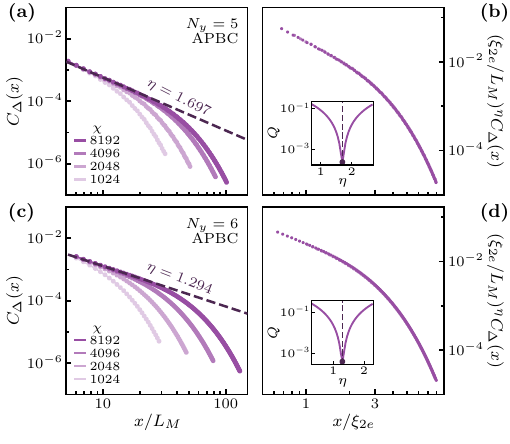}
    \caption{Power-law fitting and scaling collapse result at $N_y=5$ and $N_y=6$, yielding  $\eta=1.697$ and $\eta=1.294$, respectively. (a,c) show the Cooper-Cooper correlation function for each bond dimension, while (b,d) show the optimal scaling collapse. Insets show the loss function $Q$ as a function of $\eta$, whose minimum gives the optimal exponent.   
    Parameters are identical to Fig.~\ref{fig:2_SC_evidence}.}
    \label{fig:app:scaling_collapse}
\end{figure}

\section{Curve Fitting Details}
\label{app:curve_fitting}

\subsection{Stiffness}
The superfluid stiffness follows from the second-order change in the energy density given in Eq.~\eqref{eq:SC_stiffness}, which we reproduce here for convenience:
\begin{equation}
    \Delta \mathcal E(\delta\Phi_y)
    =
    \frac{\rho_s}{2}
    \left[
    \frac{2\delta\Phi_y}{N_y L_M}
    \right]^2,
\end{equation}
where $\delta \Phi_y = \Phi_y - \pi$. In practice, we compute the DMRG ground state at a range of $\delta \Phi_y$'s and fit them to a parabola. While large $\delta \Phi_y$ values always behave quadratically, very close to $\Phi_y =\pi$ the changes in energy can be more erratic. In some cases, this may be set by the norm error or convergence of the DMRG, but a physical origin cannot be excluded. We therefore perform a parabolic fit excluding the points closest to $\Phi_y = \pi$. At finite $u_D$, the coefficient of the linear term in $\delta \Phi_y$ becomes non-zero and is fitted as well. We estimate our fitting error between different choices of which points to include to be on the $10\%$ level at most, and usually on the $1\%$ level. A well-fit point is shown in Fig.\ref{fig:app:stiffness_fitting}

\subsection{Power Laws}

The correlation function $C_O(x;\chi) = \braket{\hat{O}(x)^\dagger \hat{O}(x)}$ for an operator $\hat{O}$ with quasi-long range order manifests in DMRG as a power law $C(x) \sim x^{-\eta}$ in the limit $\chi \to \infty$. However, at each finite $\chi$, $C(x)$ must decay exponentially at large distances. To fit $\eta$ from finite $\chi$ data, we take advantage of the scaling collapse form
\begin{equation}
    C_O(x, \xi_{\chi}) = \xi_{\chi}^{-\eta} g(x/\xi_{\chi})
\end{equation}
for some scaling function $g(u)$. Following e.g.~\cite{Sahay2024Superconductivity} we may then define a loss function for $\eta$ that is minimized when the curves from different $\chi$ are closest. To that end, define
\begin{equation}
    y_{\chi}(u;\eta) = \ln[\xi_{\chi}^\eta C_O(\xi_{\chi} u; \chi)]
\end{equation}
and put
\begin{equation}
    Q(\eta) = \frac{1}{N_{\chi} \Delta u} \int_{u_0}^{u_1} \sum_{\chi}
    \left[ y_{\chi}(u;\eta) - \overline{y}(u;\eta) \right]^2
\end{equation}
where $N_{\chi}$ is the number of bond dimensions, $\overline{y}(u;\eta) = \frac{1}{N_\chi} \sum_{\chi} y_{\chi}(u;\eta)$, $\Delta u = u_1 - u_0$ is range over which all curves are defined. As each $y_\chi(u;\eta)$ has data at different $u$'s, the data is linearly interpolated to perform the integral as a Riemann sum.  We fit using only $x > 10 L_M$ to remove the non-universal ``head'' of the curve. Fig.~\ref{fig:app:scaling_collapse} shows typical fits.

\end{document}